\pdfoutput=1
\pdfoutput=1
\pdfoutput=1
\pdfoutput=1
\pdfoutput=1
\documentclass[twocolumn,10pt]{IEEEtran}
\normalsize

\usepackage{enumerate}
\usepackage{color}

\usepackage{cite}

\usepackage{float}
\newfloat{figtab}{htb}{fgtb}
\makeatletter
\newcommand\figcaption{\def\@captype{figure}\caption}
\newcommand\tabcaption{\def\@captype{table}\caption}
\makeatother

\usepackage{graphicx}
\usepackage{booktabs}
\usepackage{algorithm}
\usepackage{bm}
\ifCLASSINFOpdf
\else
\fi

\usepackage{amsthm}
\usepackage{amssymb,amsfonts}
\usepackage{amsfonts,balance}

\usepackage[cmex10]{amsmath}

\newtheorem{theorem}{Theorem}

\usepackage{algorithmic}

\usepackage{array}
\usepackage{multirow,makecell}
\usepackage[caption=false,font=footnotesize]{subfig}

\usepackage{verbatim}

\begin{document}
%
\title{Learning-based Prediction, Rendering and Transmission for Interactive Virtual Reality in RIS-Assisted Terahertz Networks}
%
%
\author{Xiaonan~Liu,~\IEEEmembership{Student Member,~IEEE,}
        Yansha~Deng,~\IEEEmembership{Member,~IEEE,} Chong~Han,~\IEEEmembership{Member,~IEEE,}\\ and Marco Di Renzo,~\IEEEmembership{Fellow,~IEEE}\\
\thanks{X. Liu and Y. Deng are with the Department of Engineering, King’s College London, London, WC2R 2LS, U.K. (e-mail:\{xiaonan.liu, yansha.deng\}@kcl.ac.uk). (Corresponding author: Yansha Deng).}
\thanks{C. Han is with Terahertz Wireless Communications (TWC) Laboratory, Shanghai Jiao Tong University, China. (e-mail:\{chong.han\}@sjtu.edu.cn).}
\thanks{M. Di Renzo is with Universit\'e Paris-Saclay, CNRS, CentraleSup\'elec, Laboratoire des Signaux et Syst\`emes, 3 Rue Joliot-Curie, 91192 Gif-sur-Yvette, France. (marco.di-renzo@universite-paris-saclay.fr)} }

\maketitle

\begin{abstract}
The quality of experience (QoE) requirements of wireless Virtual Reality (VR) can only be satisfied with high data rate, high reliability, and low VR interaction latency. This high data rate over short transmission distances may be achieved via abundant bandwidth in the terahertz (THz) band. However, THz waves suffer from severe signal attenuation, which may be compensated by the reconfigurable intelligent surface (RIS) technology with programmable reflecting elements. Meanwhile, the low VR interaction latency may be achieved with the mobile edge computing (MEC) network architecture due to its high computation capability. Motivated by these considerations, in this paper, we propose a MEC-enabled and RIS-assisted THz VR network in an indoor scenario, by taking into account the uplink viewpoint prediction and position transmission, MEC rendering, and downlink transmission. We propose two methods, which are referred to as centralized online Gated Recurrent Unit (GRU) and distributed Federated Averaging (FedAvg), to predict the viewpoints of VR users. In the uplink, an algorithm that integrates online Long-short Term Memory (LSTM) and Convolutional Neural Networks (CNN) is deployed to predict the locations and the line-of-sight and non-line-of-sight statuses of the VR users over time. In the downlink, we further develop a constrained deep reinforcement learning algorithm to select the optimal phase shifts of the RIS under latency constraints. Simulation results show that our proposed learning architecture achieves near-optimal QoE as that of the genie-aided benchmark algorithm, and about two times improvement in QoE compared to the random phase shift selection scheme.
\end{abstract}




%
\IEEEpeerreviewmaketitle

\section{Introduction}
Wireless virtual reality (VR) can be a potential solution in breaking geographical boundaries, providing the VR users with a sense of total presence and immersion under VR interaction latency, and may unleash plenty of novel VR applications \cite{VR_system}. To achieve this vision, we still face unique challenges, including how to support real-time VR interaction under low interaction latency (in the order of tens of milliseconds), high resolution 360 degree VR video transmission under high data rates, seamless connectivity for moving VR users even under unstable wireless channels, and satisfy the asymmetric and coupled uplink and downlink requirements \cite{Hu1}.

To address the above challenges, terahertz (THz) communication can be a promising enabler for high rate, high reliability, and low VR interaction latency \cite{THz2}. However, the THz transmission suffers from severe propagation attenuation and water-molecular absorption loss because of its high frequency, which limits the propagation distance \cite{Propagation}. That is the reason why the THz communication system is usually deployed in an indoor scenario. Note that the indoor environment can be complex with physical obstacles, such as walls and furniture, which may block the line-of-sight (LoS) communication links \cite{THz3,THz5}.

To address the severe path attenuation of THz and support transmission for users in non-line-of-sight (NLoS) areas, reconfigurable intelligent surface (RIS) can be an effective approach to create a second virtual LoS path and enhance the coverage \cite{Macro1,Macro2,Macro4,Macro5,Macro6}.
The RIS is a planar surface that consists of a number of small-unit reflectors, and is equipped with a low-cost sensor and controlled with a simple processor. Each reflecting element of the RIS can reflect incident electromagnetic waves independently with an adjustable phase shift. Through deploying the RIS in the THz network and smartly adjusting the phase shift of all the elements, the THz signals between the transmitters and receivers can be reconfigured flexibly to support the THz transmission for the users in NLoS areas \cite{IRS7,Macro3,Macro7}. 

It is important to note that the VR interaction latency, which is composed of the uplink transmission latency, the rendering latency, and the downlink transmission latency, is also one of the key requirements in VR service. Violating the VR interaction latency constraint can lead to motion sickness and discomfort \cite{Hu1}. Rendering real-time high quality VR videos via a computing unit with high processing capabilities can be a  potential solution to reduce the VR interaction latency. To do so, mobile edge computing (MEC) can be introduced to shift the heavy VR video computation load from the VR device to the MEC server \cite{Jun_zhang_mec}. 

Motivated by the above, in this paper, we focus on optimizing the QoE of VR users in a MEC-enabled and RIS-assisted THz VR network in an indoor scenario, and we develop a novel learning strategy to efficiently optimize the long-term QoE in THz VR systems. The main contributions are summarized as follows:
\begin{itemize}
    \item We propose a MEC-enabled and RIS-assisted THz VR network in an indoor scenario, taking into account the uplink transmission, MEC rendering, and downlink VR video transmission.
    \item In the uplink, we use a two-ray uplink transmission to deliver the actual viewpoints or learning models to the MEC. Based on the historical and current viewpoint of the VR user from real VR datasets \cite{VRdataset}, we propose two methods, which are referred to as the centralized online Gated Recurrent Unit (GRU) algorithm and the distributed Federated Averaging (FedAvg) algorithm, to predict the dynamical viewpoint preference of the VR users over time. By doing so, the predicted field of view (FoV) can be rendered and transmitted in advance, with the aim to reduce the VR interaction latency.
    \item We also propose an algorithm that integrates online Long-Short Term Memory (LSTM) and Convolutional Neural Networks (CNN) to predict the new positions of the VR users based on their historical positions delivered via the uplink, which are then used to predict the LoS or the NLoS status of the VR users.
    \item With the predicted viewpoint and the LoS/NLoS status of each VR user as the inputs, we propose a constrained Deep Reinforcement Learning (C-DRL) algorithm to optimize the long-term QoE of VR users under VR interaction latency constraints, by selecting the optimal phase shifts of the RIS reflecting elements. Through comparison with non-learning-based methods, we show that our proposed ensemble learning architecture with GRU, LSTM, CNN, and C-DRL achieves near optimal QoE similar to that offered by an exhaustive-search algorithm, and enhances of about two times the QoE compared to the random phase shift selection scheme. 
\end{itemize}

The rest of this paper is organized as follows. Section II presents the related works. The system model and problem formulation are proposed in Section III.
The learning algorithms for THz VR systems are presented in
Section IV. The simulation results and conclusions are described in Section V and Section VI, respectively.

\section{Related Works}
In this section, related works on THz VR systems, RIS-assisted THz networks, and MEC-enabled wireless VR networks are briefly introduced in the following three subsections.

\subsubsection{Wireless VR/AR system in THz} In \cite{THz_VR1,THZ_IRS4,Yu1,THz_AR}, the reliability, transmission rate, viewpoint rendering, and energy consumption of a THz VR/AR system were investigated. Specifically, in \cite{THz_VR1}, the reliability of VR services in the THz band was studied, and the theoretical analysis of the end-to-end (E2E) delay was performed. In \cite{THZ_IRS4}, a risk-based framework was proposed to optimize the rate and reliability of THz-band for VR applications. In \cite{Yu1}, through jointly optimizing the viewport rendering offloading and downlink transmit power control, the long-term energy consumption of a THz wireless access-based MEC system for high quality immersive VR video services was minimized. In \cite{THz_AR}, the age
of information (AoI) of AR services in a THz cellular network with RISs was studied, and the cumulative distribution function (CDF) of the AoI was derived for two different scheduling policies, which are the last come first served (LCFS) queues and the first come first served (FCFS) queues. However, in \cite{THz_VR1}, \cite{THZ_IRS4}, and \cite{THz_AR}, the authors mainly focused on the theoretical analysis of the reliability and E2E delay of the THz VR/AR. In \cite{Yu1}, the authors mainly optimized the long-term energy consumption via deep reinforcement learning, rather than the QoE and VR interaction latency, without the consideration of RIS.

\subsubsection{RIS-assisted THz network} In \cite{THZ_IRS2,Nano,C_Huang,li_wei,Wu1,wU2,Chaccour1,Chaccour2,Chaccour3}, the sum rate maximization problem in RIS-assisted THz network was investigated. Specifically, in \cite{THZ_IRS2}, based on channel estimation, a deep neural network (DNN) was proposed to select the optimal phase shift configurations to maximize the sum rate in the RIS-assisted THz multiple-input multiple-output (MIMO) system. In \cite{Nano}, through optimizing the beamforming vector of the transmitter and reflection coefficients matrix of the RIS, the coverage probability in indoor THz communication scenarios was improved, and a low complexity phase shift search scheme was used to achieve near-optimal coverage performance. In \cite{C_Huang}, energy-efficient designs for both the transmit power allocation and the phase shifts of the RIS subject to downlink multi-user communication were developed. In \cite{li_wei}, the downlink of an RIS-empowered multiple-input single-output (MISO) communication system was considered, where the alternating least squares and vector approximate message passing methods were used to estimate channels between the base station (BS) and the RIS, as well as the channels between the RIS and users. In \cite{Wu1}, RIS-assisted secure wireless communications were investigated, and the successive convex approximation-based algorithm was used to solve the transmit covariance matrix optimization problem, which maximized the secrecy rate. In \cite{wU2}, the network architecture and spectrum access of AI-enabled Internet of Things in 5G and 6G networks were proposed. In \cite{Chaccour1}, a comprehensive roadmap outlining the seven defining features of THz wireless systems that guarantee a successful deployment in future wireless generations was proposed. In \cite{Chaccour2} and \cite{Chaccour3}, a novel hybrid beamforming scheme for multi-hop RIS-assisted communication networks was proposed to improve the coverage range of THz networks. Nevertheless, in \cite{THZ_IRS2,Nano,C_Huang,li_wei,Wu1,wU2,Chaccour1,Chaccour2,Chaccour3}, the authors mainly optimized the phase shift and beamforming vector to maximize the sum rate of RIS-assisted THz networks using a DNN or non-learning based approaches.

\subsubsection{MEC-enabled wireless VR network}
In \cite{VR6} and \cite{peng1}, a joint caching and computing optimization problem was formulated to minimize the
average transmission rate and maximize the average tolerant delay, respectively. In \cite{Cui1}, through effectively exploiting the characteristics of multi-quality tiled 360 VR videos and computation resources, the optimal wireless streaming of a multi-quality tiled 360 VR video to multiple users in wireless networks was investigated to decrease the energy consumption. In \cite{xiaonan}, a decoupled learning strategy was developed to optimize the real-time VR video streaming in wireless networks, by taking into account the FoV prediction and rendering MEC association. Through the design of centralized/distributed deep reinforcement learning algorithms to select proper MECs to render and transmit predicted VR video frames, the QoE was enhanced and the VR interaction latency was reduced. However, the authors of \cite{VR6,peng1,Cui1} mainly used convex optimization to optimize the wireless VR network, which can not guarantee the long-term QoE of the VR users. The authors of \cite{xiaonan} only considered the MEC association problem without using real VR datasets.

\begin{figure}[!t]
    \centering
    \includegraphics[width=3.5 in]{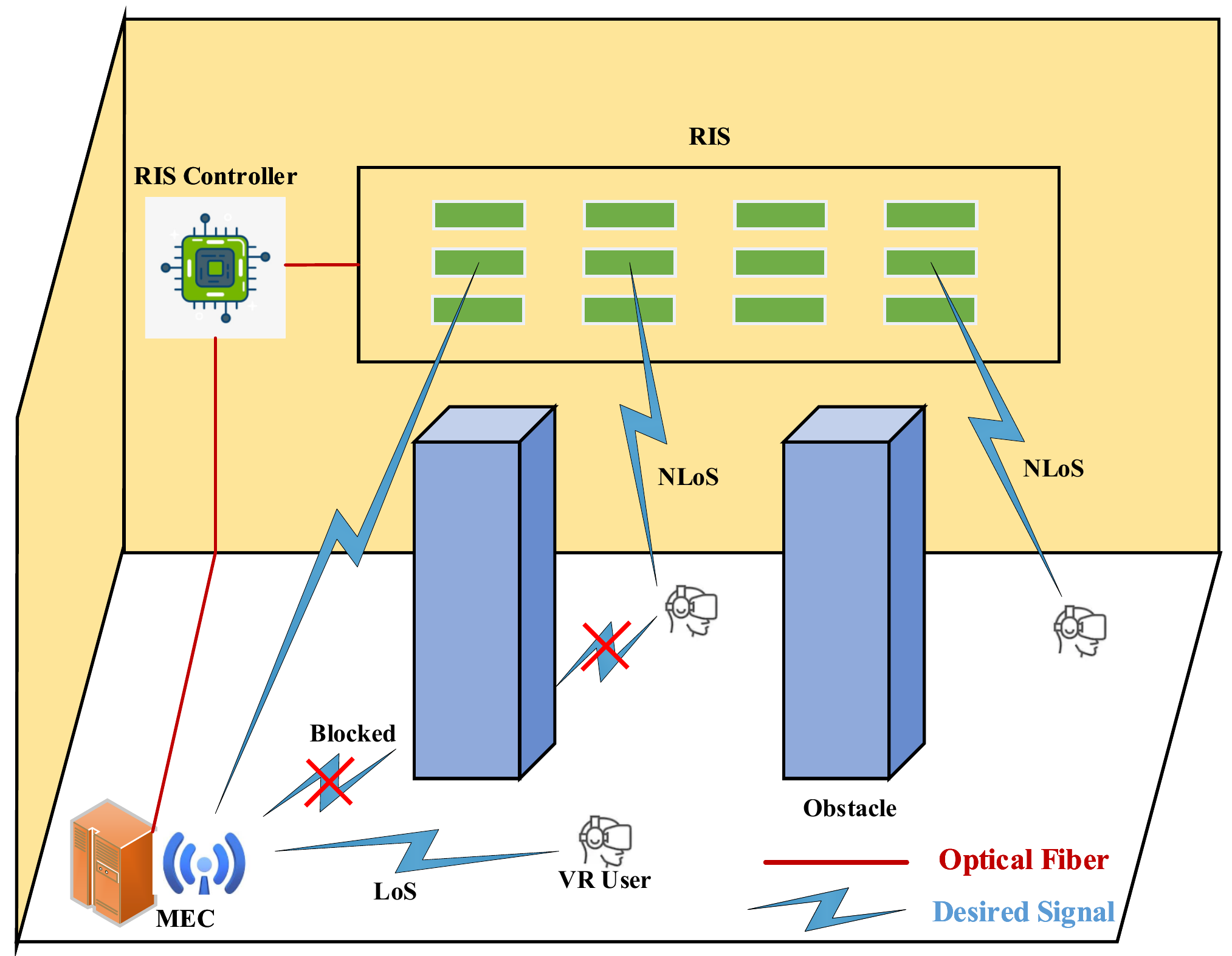}
    \caption{Wireless VR system in THz network.}
    \label{basic_modules}
\end{figure}

\section{System Model and Problem Formulation}
We consider an indoor scenario, where an RIS that comprises $N$ reflecting elements is deployed to assist the uplink and downlink transmission between a MEC and $K^{\text{VR}}$ VR users, as shown in Fig. 1. The MEC operating over THz frequency is equipped with $M$ antennas\footnote{Physically, the MEC and SBS are co-located in one location.} and each VR user is equipped with a single antenna, respectively. The indoor scenario is assumed to be a square with length $W$ of each side. The RIS is connected to a smart controller that communicates with the MEC via a wired link for cooperative transmission and information exchange, such as channel state information (CSI), and phase shifts control of all reflecting elements \cite{IRS5}. Due to the substantial path loss in THz transmission, we only consider  the THz signal reflected by the RIS for the first time and ignore the signals that are reflected for twice or more times following \cite{IRS6}. 

\subsection{VR User Mobility}
We present a mobility model based on the VR user movements in the indoor VR scenario, which is the so-called virtual reality mobility model (VRMM) \cite{VRMM}. The VRMM includes the following parameters: start location, destination location, speed, and moving direction. We assume that there are four directions for the VR user to select, namely, up, down, left, and right. We split the indoor area into $W\times W$ grids. When the VR user is at the start location, it sets its destination location, speed, and moving direction, and transmits its current location at each time slot to the MEC server through uplink transmission. Note that the location of the VR user for the next time slot is determined by the location of the current time slot rather than the locations in previous time slots, so that the mobility of the VR user in the indoor area follows the Markov property. When the VR user arrives at the destination location, it sets a new destination location and moves forward to it with a given speed.

\begin{figure}[!t]
    \centering
    \includegraphics[width=3.0 in]{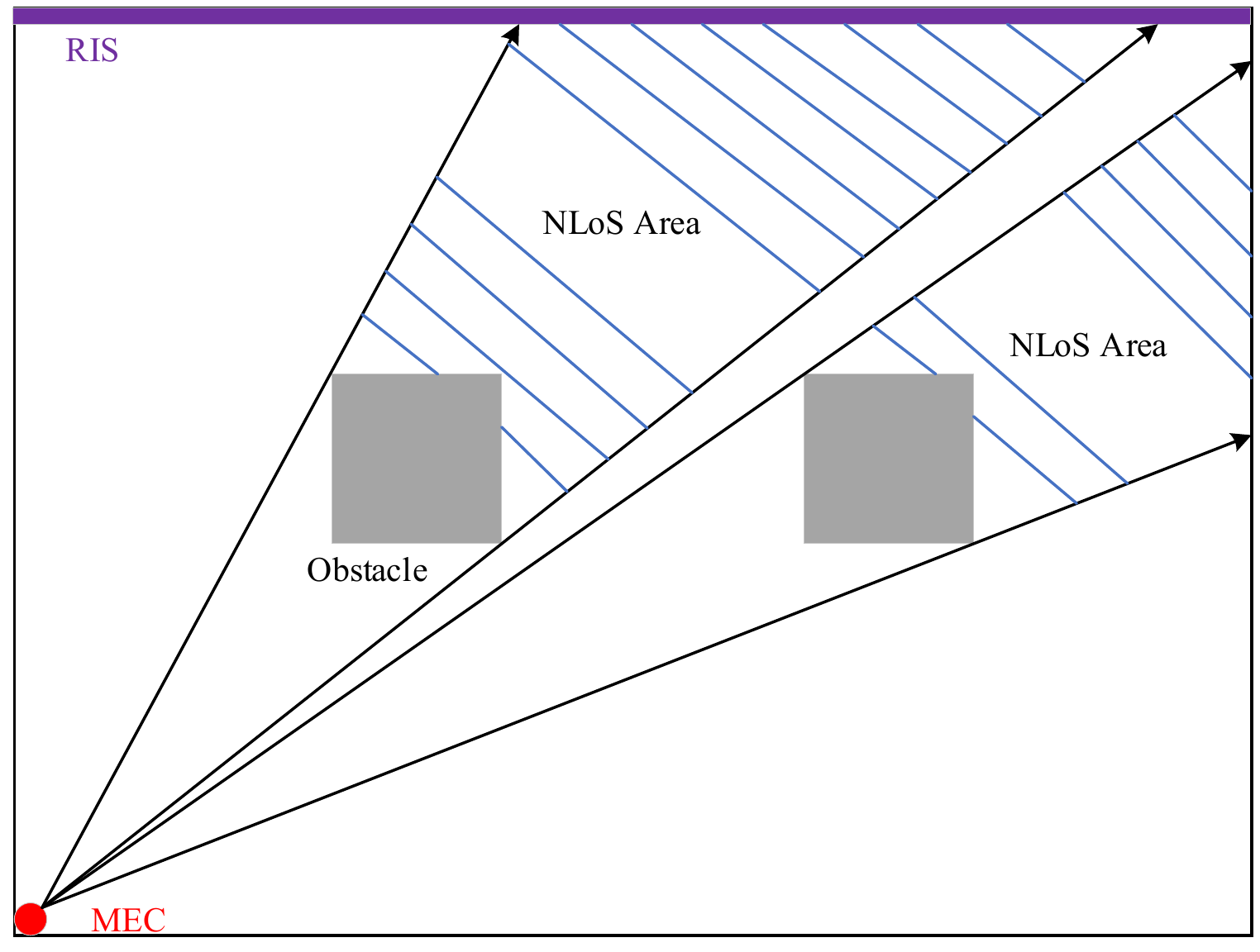}
    \caption{Illustration of THz network in the presence of obstacles.}
    \label{basic_modules}
\end{figure}

\subsection{Indoor Blockage}
Due to the severe signal attenuation and narrow wave spread in THz frequencies, the THz transmission is very sensitive to the presence of obstacles \cite{THZ1}.  When the VR users are moving in an indoor scenario, the blockage between the MEC and the $k$th VR user can be caused by an obstacle and the other VR users with higher heights that are located near to the $k$th VR user. For simplicity, we map the 3D indoor scenario into a 2D image. In Fig. 2, when VR users are behind the obstacle, they are directly blocked by the obstacle. As shown in Fig. 3, we assume that the height of the MEC is $h_A$, the height of the VR user 2 is $h_B$ $(h_B < h_A)$, the height of the VR user 1 is $h_U$ $(h_U < h_B)$, the distance between the VR user 2 with height $h_B$ and the MEC is $l$, and the distance between the VR user 1 with height $h_U$ and the MEC is $x$.

\begin{figure}[!t]
    \centering
    \includegraphics[width=3.5 in]{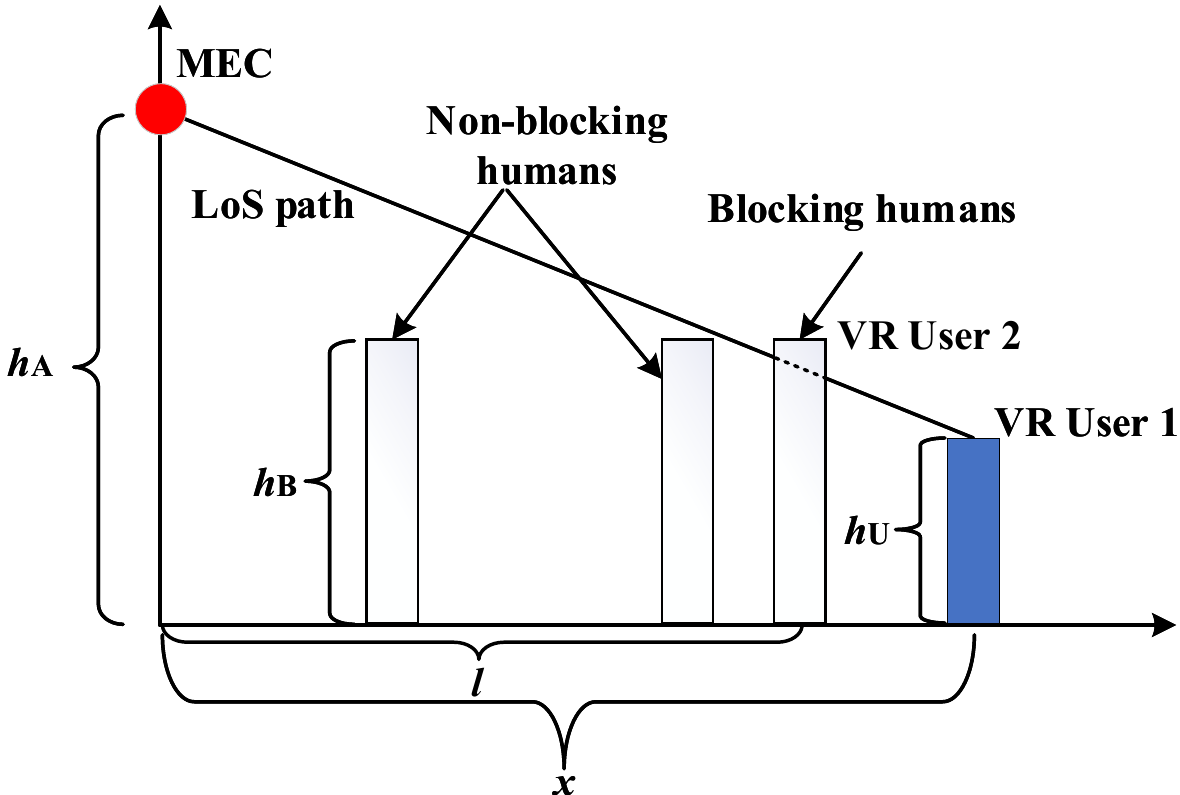}
    \caption{Illustration of a single THz transmission link in the presence of blocker via other VR user with higher height.}
    \label{basic_modules}
\end{figure}

$\textbf{Definition 1:}$ When the MEC server, the VR user 2, and the VR user 1 are located in the same line in the 2D plane, the VR user 1 will be blocked by the VR user 2 if its distance in the 2D plane is less than $\frac{(h_A - h_U)l}{h_A-h_B}$.
\begin{proof}
According to Fig. 3, the coordinates of the MEC, the VR user 2, and the VR user 1 in the 2D plane are denoted as $(0, h_A)$, $(l, h_B)$, and $(x, h_U)$, respectively. Applying slope-intercept equation, we can calculate the line equation across the point $(0, h_A)$ and $(l, h_B)$ as
\begin{equation}
    y = \frac{h_B - h_A}{l}x + h_A.
\end{equation}
For $y = h_U$, $x$ is computed as $\frac{(h_A - h_U)l}{h_A-h_B}$. Thus, the distance between the VR user 2 and the VR user 1 is computed as $\frac{(h_B - h_U)l}{h_A-h_B}$.
\end{proof}

Due to the blockage caused by the obstacles, such as pillars, walls, or other VR users, the THz transmission between the MEC server and the VR users can be enhanced by the RIS, where each passive reflecting element can change the phase shift of the THz wave \cite{Nano}. In our model, we define the MEC-VR user link as the line-of-sight (LoS) link, and the MEC-RIS-VR user link as non-LoS (NLoS) link. It is important to note that through obtaining the current and historical locations and LoS/NLoS statuses of the VR users, the MEC server can predict the LoS/NLoS statuses of the VR users at each time slot.

\subsection{THz Uplink Transmission}
At the start of each time slot, the VR user transmits its actual viewpoint and location to the MEC via uplink transmission. Because of the mobility of the VR user, it may enter the LoS or NLoS region. To guarantee the reliability of the uplink transmission, we consider a two-ray uplink transmission. One ray is the LoS link, and the other is the NLoS link. For the VR user in the LoS region, the received signals include the ones from LoS and NLoS links. While for the VR user in the NLoS region, the received signal only includes the signal from the NLoS link. For the $k$th VR user, the transmitted two-ray signals through the uplink transmission at the $t$th time slot are denoted as
\begin{align}
    \textbf{y}_{k}^{\rm{up}}(t) \! = &\textbf{u}_{k}^{H}(t){\textbf{h}}_{k}^{\rm{up}}(t){x}_{k}^{\rm{up}}(t) +
     \\&{\textbf{u}}_{k}^{H}(t)\textbf{G}^{\rm{up}}(t)\boldsymbol{\Theta}^{\rm{up}}(t)\textbf{g}_{k}^{\rm{up}}(t){x}_{k}^{\rm{up}}(t) + 
    \nonumber\\&\sum_{i =1, i\neq k}^{K^{\rm{VR}}}\!\!\!\!\textbf{u}_{k}^{H}(t){\textbf{h}}_{i}^{\rm{up}}(t){x}_{i}^{\rm{up}}(t) + 
    \nonumber\\&\sum\limits_{i =1, i\neq k}^{K^{\rm{VR}}}\!\!\!\!{\textbf{u}}_{k}^{H}(t)\textbf{G}^{\rm{up}}(t)\boldsymbol{\Theta}^{\rm{up}}(t)\textbf{g}_{i}^{\rm{up}}(t)x_{i}^{\rm{up}}(t) + \textbf{n}_{k}^{\rm{up}}(t),\nonumber
\end{align}
where $\textbf{u}_{k}^{H}(t)\in\mathbb{C}^{1\times M}$ is the beamforming vector of the $k$th VR user at the $t$th time slot, which can be denoted as $\frac{{\textbf{h}}_{k}^{\rm{up}}(t) + \textbf{G}^{\rm{up}}(t)\boldsymbol{\Theta}^{\rm{up}}(t)\textbf{g}_{k}^{\rm{up}}(t)}{\|{\textbf{h}}_{k}^{\rm{up}}(t) + \textbf{G}^{\rm{up}}(t)\boldsymbol{\Theta}^{\rm{up}}(t)\textbf{g}_{k}^{\rm{up}}(t)\|}$ \cite{IRS4}. In (2), ${\textbf{h}}^{\rm{up}}(t)\in\mathbb{C}^{M\times 1}$ is the channel vector between the MEC and the $k$th VR user at the $t$th time slot, ${x}_{k}^{\rm{up}}(t)$ is the transmitted data symbol of the $k$th VR user, and is set as discrete random variable with zero mean and unit variance, $\textbf{g}_{k}^{\rm{up}}(t)\in\mathbb{C}^{N\times 1}$ is the channel matrix between the $k$th VR user and the RIS, $\textbf{G}^{\rm{up}}(t)\in\mathbb{C}^{M\times N}$ is the channel matrix between the RIS and the MEC, and $\textbf{n}_{k}^{\rm{up}}(t)$ is the additive white Gaussian noise of the $k$th VR user with zero mean and $\hat{\sigma}_{k}^2$ variance. Meanwhile, $\sum_{i =1, i\neq k}^{K^{\rm{VR}}}\textbf{u}_{k}^{H}(t){\textbf{h}}_{i}^{\rm{up}}(t){x}_{i}^{\rm{up}}(t)$ and $\sum_{i =1, i\neq k}^{K^{\rm{VR}}}{\textbf{u}}_{k}^{H}(t)\textbf{G}^{\rm{up}}(t)\boldsymbol{\Theta}^{\rm{up}}(t)\textbf{g}_{i}^{\rm{up}}(t)x_{i}^{\rm{up}}(t)$ are the interferences from the LoS and NLoS links of other VR users, respectively. Let $\boldsymbol{\theta} = [\theta_{1},...,\theta_{N}]$ denote the selected phase shift set of $N$ reflection elements, where $\theta_{n}\in[0,2\pi]$ denotes the phase shift of the $n$th reflecting element of the RIS, which can be carefully adjusted by an RIS controller. Here, the reflection coefficients matrix $\boldsymbol{\Theta}^{\rm{up}}(t)$ is presented as
\begin{equation}
\boldsymbol{\Theta}^{\text{up}}(t) = \text{diag}({e^{j\theta_{1}^{\text{up}}}(t),...,e^{j\theta_{N}^{\text{up}}}(t)}).
\end{equation}
For practical implementation, we assume that the phase shift of each element of the RIS can only take a finite number of discrete values. We set $b$ as the number of bits used to indicate the number of phase shift levels $\hat{L}$, where $\hat{L} = 2^b$. For simplicity, we assume that such discrete phase-shift values can be obtained by uniformly quantizing the interval $[0, 2\pi)$. Thus, the set of discrete phase shift values at each element is given by
\begin{equation}
    \mathcal{F} = \{0, \triangle\theta,..., (\hat{L}-1)\triangle\theta\},
\end{equation}
where $\triangle\theta = 2\pi/\hat{L}$ \cite{IRS6}. Now, the uplink transmission rate of the $k$th VR user at the $t$th time slot is calculated as
\begin{equation}
    R_{k}^{\rm{up}}(t)\!\! = \!\! \log_2\left|\textbf{I}\!\! + \!\! \frac{|\textbf{u}_{k}^{H}(t)({\textbf{h}}_{k}^{\rm{up}}(t) + \textbf{G}^{\rm{up}}(t)\boldsymbol{\Theta}^{\rm{up}}(t)\textbf{g}_{k}^{\rm{up}}(t)) |^2}{\textbf{I}_{k}^{\rm{up}}(t) + \hat{\sigma}_{k}^{2}\textbf{I}_{M}}\right|,
\end{equation}
where $\textbf{I}_M$ is the identity matrix, and
\begin{equation}
    \textbf{I}_{k}^{\rm{up}}(t) \!\!= \!\!\!\!\sum_{i =1, i\neq k}^{K^{\rm{VR}}}|\textbf{u}_{k}^{H}(t)({\textbf{h}}_{i}^{\rm{up}}(t) + \textbf{G}^{\rm{up}}(t)\boldsymbol{\Theta}^{\rm{up}}(t)\textbf{g}_{i}^{\rm{up}}(t))|^2.
\end{equation}
According to (5), the uplink transmission rate of the VR user in the LoS area is determined by both the LoS and NLoS links. The uplink transmission rate of the $k$th VR user in the NLoS area is only affected by the NLoS link, and $\textbf{u}_{k}^{H}(t){\textbf{h}}_{k}^{\rm{up}}(t) = 0$.

\subsection{VR Viewpoint Prediction}
By predicting the viewpoint preference of the VR user with centralized online GRU algorithm or distributed FedAvg algorithm, the corresponding FoV can be rendered and transmitted in advance in order to decrease the VR interaction latency. When the VR user watches the VR video frames, the viewpoint is determined by three degrees of freedom, which are $X$, $Y$, and $Z$ axes. Thus, predicting the viewpoint of the VR user can be converted to predicting the $X$, $Y$, and $Z$ angles. We consider a sliding window to predict the viewpoint of the VR user in continuous time slots. The future viewpoint of the VR user can be predicted according to the current and past rotation statuses. To guarantee the prediction accuracy, we use the online GRU algorithm to predict the viewpoint of the VR user at each time slot. Specifically, we use Mean Square Error (MSE) as the cost function in each training step to update the parameters of the online GRU model, which is calculated as
\begin{equation}
    \text{MSE}_{t}^{k} = \frac{1}{K^{\rm{VR}}}\sum\limits_{k=1}^{K^{\rm{VR}}}(\hat{V}_{t}^{k}-V_{t}^{k})^2,
\end{equation}
where $\hat{V}_t^{k} = (\hat{X}_t^{k}, \hat{Y}_t^{k}, \hat{Z}_t^{k})$ and ${V}_t^{k} = ({X}_t^{k}, {Y}_t^{k}, {Z}_t^{k})$ are the predicted and actual viewpoint of the $k$th VR user at the $t$th time slot, respectively.
At the $(t-1)$th time slot, the MEC or the VR device will predict the viewpoint $\hat{V}_t^{k}$ of the $k$th VR user for the $t$th time slot. Then, the $k$th VR user will transmit its actual viewpoint ${V}_t^{k}$ or the learning model to the MEC via uplink transmission. By comparing it with the predicted viewpoint, the parameters of the online GRU algorithm are updated, which can further improve the prediction accuracy.

\subsection{MEC Rendering}
When the VR users enjoy the VR video frames, the corresponding portion of the sphere is rendered at the MEC based on the predicted viewpoint. Through equirectangular projection (ERP) mapping \cite{ERP}, a stitched 2D image with RGB color model is rendered into the required FoV. We assume that the resolution of the FoV is $N_p\times N_v$, and the size of each pixel is $8$ bits. The size of the FoV in RGB model is calculated as
\begin{equation}
    C = 3 \times 8 \times N_p \times N_v \times V,
\end{equation}
where $3$ represents the red, green, and blue color in RGB model, and $V = 2$ is the number of viewpoints for two eyes. We assume that the execution ability of the GPU of the MEC is $F_{\text{MEC}}$, and the number of cycles required for processing one bit of input data in the MEC is $f_{\text{MEC}}$. The MEC rendering latency is calculated as
\begin{equation}
    T_{\text{render}} = \frac{f_{\text{MEC}}C}{F_{\text{MEC}}}.
\end{equation}
From (9), we can obtain that the rendering latency for all VR users is the same.

\subsection{THz Downlink Transmission}
In the THz downlink transmission, it is possible that VR users may be blocked by the obstacles or VR users with higher heights, as shown in Fig. 2 and Fig. 3. For the VR users that are not blocked by obstacles and other VR users, the MEC directly performs transmission in LoS channel, otherwise, it is served by NLoS channel aided by the RIS \cite{IRS1,IRS2,IRS3}.

We consider a multi-input single-output (MISO) THz channel. We use the sets $\mathcal{V}_{\rm{LoS}}$ and $\mathcal{V}_{\rm{NLoS}}$ to denote the LoS and NLoS VR user groups, respectively. For the $k$th VR user in the LoS group, the received signal from the MEC at the $t$th time slot is denoted as
\begin{align}
    \textbf{y}_{k}^{\rm{LoS}}(t) &= \textbf{h}_{k}^{H}(t)\textbf{v}_{k}^{\rm{LoS}}(t)x_{k}^{\rm{LoS}}(t) + \nonumber \\
    &\sum\limits_{i\neq k, i\in\mathcal{V}_{\rm{LoS}} }\textbf{h}_{k}^{H}(t)\textbf{v}_{i}^{\rm{LoS}}(t)x_{i}^{\rm{LoS}}(t)+
    \nonumber \\
    &\sum\limits_{j\in\mathcal{V}_{\rm{NLoS}}}\textbf{h}_{k}^{H}(t)\textbf{v}_{j}^{\rm{NLoS}}(t)x_{j}^{\rm{NLoS}}(t) + \textbf{n}_{k}(t),
\end{align}
where $\textbf{h}_{k}(t)\in\mathbb{C}^{M\times 1}$ is the channel vector between the MEC and the $k$th VR user, $\textbf{v}_{k}^{\rm{LoS}}(t)\in\mathbb{C}^{M\times 1}$ and $\textbf{v}_{j}^{\rm{NLoS}}(t)\in\mathbb{C}^{M\times 1}$ are the beamforming vectors of the $k$th VR user in the LoS group, and the $j$th VR user in the NLoS group, respectively. In (10), $\textbf{v}_{k}^{\rm{LoS}}(t)$ can be denoted as $\frac{\textbf{h}_{k}(t)}{\|\textbf{h}_{k}(t)\|}$, and $x_{k}^{\rm{LoS}}(t)$ and $x_{j}^{\rm{NLoS}}(t)$ indicate the transmitted data symbol for the $k$th VR user in the LoS group and the $j$th VR user in the NLoS group, respectively, and are defined as discrete random variable with zero mean and unit variance. We assume that $x_{k}^{\rm{LoS}}(t)$ and $x_{j}^{\rm{NLoS}}(t)$ are independent from each other. Meanwhile, $\sum_{i\neq k, i\in\mathcal{V}_{\rm{LoS}}}\textbf{h}_{k}^{H}(t)\textbf{v}_{i}^{\rm{LoS}}(t)x_{i}^{\rm{LoS}}(t)$ and $\sum_{j\in\mathcal{V}_{\rm{NLoS}}}\textbf{h}_{k}^{H}(t)\textbf{v}_{j}^{\rm{NLoS}}(t)x_{j}^{\rm{NLoS}}(t)$
are the interference from the MEC. In addition, $\textbf{n}_{k}(t)\sim\mathcal{CN}(0, \sigma_{k}^{2}\textbf{I}_{M})$ is the additive white Gaussian noise at the $k$th VR user in the LoS group. The transmission rate between the MEC and the $k$th VR user in the LoS group at the $t$th time slot is expressed as
\begin{equation}
    R_{k}^{\rm{LoS}}(t) = \log_{2}\left(1+\frac{|\textbf{h}_{k}^{H}(t)\textbf{v}_{k}^{\rm{LoS}}(t)|^2}{\textbf{I}_{k}^{\rm{LoS}}(t) + \sigma_{k}^{2}}\right),
\end{equation}
where
\begin{equation}
    \textbf{I}_{k}^{\rm{LoS}}(t) \!\!=\!\!\!\!\!\!\!\!\! \sum\limits_{i\in\mathcal{V}_{\rm{LoS}}, i\neq k}\!\!\!\!\!\!\!\!\!|\textbf{h}_{k}^{H}(t)\textbf{v}_{i}^{\rm{LoS}}(t)|^2 + \!\!\!\!\!\!\sum\limits_{j\in\mathcal{V}_{\rm{NLoS}}}\!\!\!\!\!\!|\textbf{h}_{k}^{H}(t)\textbf{v}_{j}^{\rm{NLoS}}(t)|^2.
\end{equation}

For the VR users in the NLoS group, the signal between the MEC and the $b$th VR user at the $t$th time slot is presented as
\begin{align}
    &\textbf{y}_{b}^{\rm{NLoS}}(t) = \textbf{g}_{b}^{H}(t)\boldsymbol{\Theta}^{\text{down}}(t)\textbf{G}^{\text{down}}(t)\textbf{v}_{b}^{\rm{NLoS}}(t)x_{b}^{\rm{NLoS}}(t) + \nonumber \\
    &\sum\limits_{j\neq  b, j\in\mathcal{V}_{\rm{NLoS}}}\textbf{g}_{b}^{H}(t)\boldsymbol{\Theta}^{\text{down}}(t)\textbf{G}^{\text{down}}(t)\textbf{v}_{j}^{\rm{NLoS}}(t)x_{j}^{\rm{NLoS}}(t) + \textbf{n}_{b}(t),
\end{align}
where $\textbf{G}^{\text{down}}(t)\in\mathbb{C}^{N\times M}$ is the channel matrix between the MEC and the RIS, $\textbf{g}_{b}(t)\in\mathbb{C}^{N\times 1}$ is the channel matrix between the RIS and the $b$th VR user, $\textbf{v}_{b}^{\rm{NLoS}}(t)\in\mathbb{C}^{M\times 1}$ is the precoding matrix for the $b$th VR user in the NLoS group, which can be written as $\frac{\textbf{G}^{\rm{down}}(t)^{H}\boldsymbol{\Theta}^{\rm{down}}(t)\textbf{g}_{k}(t)}{\| \textbf{G}^{\rm{down}}(t)^{H}\boldsymbol{\Theta}^{\rm{down}}(t)\textbf{g}_{k}(t)\|}$, $x_{b}^{\rm{NLoS}}(t)$ is the transmitted data for the $b$th VR user, $\boldsymbol{\Theta}^{\text{down}}(t)$ is the reflection coefficients matrix of the RIS. Note that $\boldsymbol{\Theta}^{\text{down}}(t)$ is written as 
\begin{equation}
\boldsymbol{\Theta}^{\text{down}}(t) = \text{diag}({e^{j\theta_{1}^{\text{down}}}(t),...,e^{j\theta_{N}^{\text{down}}}(t)}),
\end{equation}
and $\textbf{n}_{b}(t)$ is the additive white Gaussian noise of the $b$th VR user with zero mean and $\sigma_k^2$ variance. $\sum_{j\neq  b, j\in\mathcal{V}_{\rm{NLoS}}}\textbf{g}_{b}^{H}(t)\boldsymbol{\Theta}^{\text{down}}(t)\textbf{G}^{\text{down}}(t)\textbf{v}_{j}^{\rm{NLoS}}(t)x_{j}^{\rm{NLoS}}(t)$ is the interference from the RIS. Then, the downlink transmission rate of the $b$th VR user in the NLoS group is written as
\begin{equation}
    R_{b}^{\rm{NLoS}}(t) = \log_2\left(1+ \frac{|\textbf{g}_{b}^{H}(t)\boldsymbol{\Theta}^{\text{down}}(t)\textbf{G}^{\text{down}}(t)\textbf{v}_{b}^{\rm{NLoS}}(t)|^2}{\textbf{I}_{b}^{\rm{NLoS}}(t) + \sigma_{b}^2}\right),
\end{equation}
where
\begin{equation}
    \textbf{I}_{b}^{\rm{NLoS}}(t) = \!\!\!\!\!\!\sum\limits_{j\neq  b, j\in\mathcal{V}_{\rm{NLoS}}}\!\!\!\!\!\!|\textbf{g}_{b}^{H}(t)\boldsymbol{\Theta}^{\text{down}}(t)\textbf{G}^{\text{down}}(t)\textbf{v}_{j}^{\rm{NLoS}}(t)|^2.
\end{equation}

\begin{figure}[!t]
    \centering
    \includegraphics[width=3.0 in]{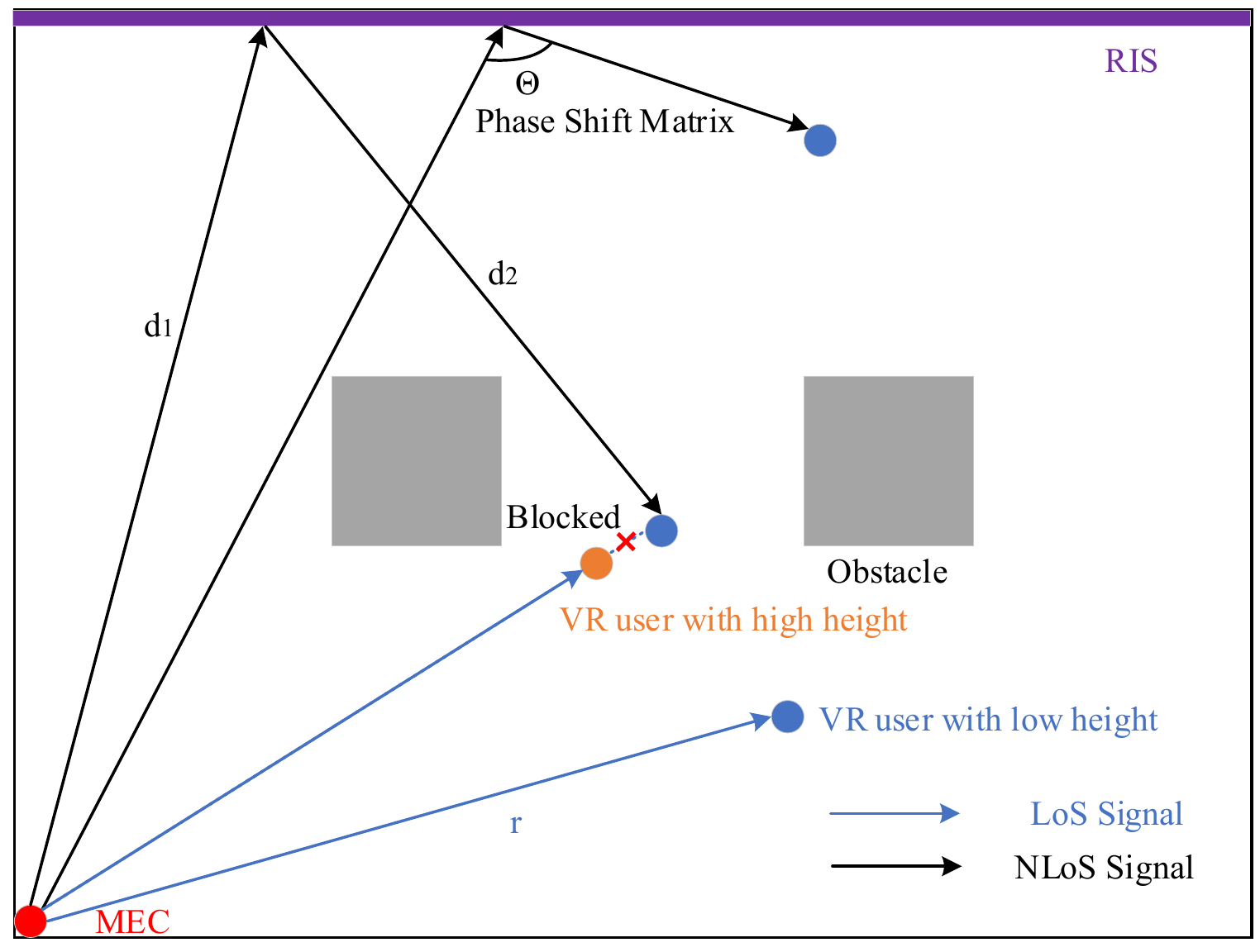}
    \caption{Illustration of THz channel model in the presence of obstacles.}
    \label{basic_modules}
\end{figure}

\subsection{THz Channel Model}
The THz channel model in the presence of obstacles is shown in Fig. 4. In THz communication, the power of the scattering component is generally much lower than that of LoS component. Thus, we ignore the scattering component, and the LoS channel is expressed as
\begin{equation}
    \tilde{\textbf{h}}_{k}(t) = h_{f,d_{k}}^{\rm{LoS}}(t)\textbf{a}_{k,\phi_{k}}^{\rm{LoS}}(t),
\end{equation}
where $\tilde{\textbf{h}}_{k}(t)=\{{\textbf{h}}_{k}^{\rm{up}}(t), \textbf{h}_{k}(t)\}$, LoS channel function $h_{\rm{LoS}}(f,d_{k})$ consists of a spreading loss function and a molecular absorption loss function, which is presented as
\begin{equation}
    h_{f,d_{k}}^{\rm{LoS}}(t) = \frac{c}{4\pi fd_{k}}e^{-\frac{\tau(f)d_{k}}{2}}e^{-j2\pi f\delta_{\rm{LoS},k}(t)},
\end{equation}
where $c$ is the speed of light. Assuming that the RIS can be installed on the wall or ceiling of the indoor scenario with height $H$, the location of the reflecting unit can be presented as $L_{\rm{RIS}} = [X_{\rm{RIS}}, Y_{\rm{RIS}}, H_{\rm{RIS}}]$. The location of the MEC can be denoted as $L_{\rm{MEC}} = [X_{\rm{MEC}}, Y_{\rm{MEC}}, H_{\rm{MEC}}]$. The location of the $k$th VR user can be written as $L_{k} = [X_{k}, Y_{k}, H_{k}]$.  The distance between the MEC and the $k$th VR user is denoted as $d_{k}$, which is calculated as
\begin{equation}
    d_{k} = \sqrt{(X_{\rm{MEC}} - X_{k})^2 + (Y_{\rm{MEC}} - Y_{k})^2 + (H_{\rm{MEC}} - H_{k})^2},
\end{equation}
$f$ is the carrier frequency, and $\delta_{\rm{LoS},k}(t)=\frac{d_{k}}{c}$ is the time-of-arrival of the LoS propagation of the $k$th VR user. $\tau(f)$ is the frequency-dependent medium absorption coefficient that depends on the molecular composition of the transmission medium, namely, the type and concentration of molecules found in the channel as defined in \cite{Multi-Ray}. In addition, $\textbf{a}_{k,\phi_{k}}^{\rm{LoS}}(t)$ is the normalized antenna array response vector at the MEC with $M$ antenna elements, which is written as
\begin{equation}
    \textbf{a}_{k,\phi_{k}}^{\rm{LoS}}(t) = \frac{1}{\sqrt{M}}[1, e^{j\frac{2\pi}{\lambda}\sin{(\phi_{k})}},...,e^{j\frac{2\pi}{\lambda}(M-1)\sin{(\phi_{k})}}]^{H},
\end{equation}
where $M$ is the number of antennas equipped in the MEC, $\lambda$ is the wavelength, and $\phi_{k}$ denotes the angles of departure/arrival (AoD/AoA).

For the NLoS transmission, the THz channels between the MEC and the RIS are denoted as
\begin{equation}
    \textbf{G}^{\text{up}}(t) = \eta G_{f,d_{\rm{M}-\rm{I}}}^{\rm{NLoS}}(t)\textbf{a}_{\phi_{\rm{MEC}}}^{\rm{NLoS}}(t)\textbf{a}_{\phi_{\rm{RIS}}}^{\rm{NLoS}}(t)^{H},
\end{equation}
and
\begin{equation}
    \textbf{G}^{\text{down}}(t) = \eta G_{f,d_{\rm{M}-\rm{I}}}^{\rm{NLoS}}(t)\textbf{a}_{\phi_{\rm{RIS}}}^{\rm{NLoS}}(t)\textbf{a}_{\phi_{\rm{MEC}}}^{\rm{NLoS}}(t)^{H},
\end{equation}
where $\eta$ is the path-loss compensation factor written as
\begin{equation}
    \eta = \frac{2\sqrt{\pi}fG_{\text{RIS}}N}{c}.
\end{equation}
In (23), $N$ is the number of elements on the RIS, and $G_{\text{RIS}}$ is the RIS element gain. The channel function $G_{f,d_{\rm{M}-\rm{I}}}^{\rm{NLoS}}(t)$ is written as
\begin{equation}
    G_{f,d_{\rm{M}-\rm{I}}}^{\rm{NLoS}}(t) = \frac{c}{4\pi fd_{\rm{M}-\rm{I}}}e^{-\frac{\tau(f)d_{\rm{M}-\rm{I}}}{2}}e^{-j2\pi f\delta_{\text{NLoS},\text{M-I}}(t)},
\end{equation}
where $d_{\rm{M}-\rm{I}}$ is the distance between the MEC and the RIS, $\delta_{\text{NLoS},\text{M-I}}(t) = \frac{d_{\rm{M}-\rm{I}}}{c}$ is the time-of-arrival of the NLoS propagation between the MEC and the RIS. The normalized antenna array response vectors $\textbf{a}_{\phi_{\rm{RIS}}}^{\rm{NLoS}}(t)$ of the RIS and $\textbf{a}_{\phi_{\rm{MEC}}}^{\rm{NLoS}}(t)$ of the MEC are written as
\begin{equation}
    \textbf{a}_{\phi_{\rm{RIS}}}^{\rm{NLoS}}(t) = \frac{1}{\sqrt{N}}[1, e^{j\frac{2\pi}{\lambda}\sin{(\phi_{\rm{RIS}})}},...,e^{j\frac{2\pi}{\lambda}(N-1)\sin{(\phi_{\rm{RIS}})}}]^{H},
\end{equation}
and
\begin{equation}
    \textbf{a}_{k,\phi_{\rm{MEC}}}^{\rm{NLoS}}\!(t) \!\!=\!\! \frac{1}{\sqrt{M}}[1, e^{j\frac{2\pi}{\lambda}\sin{(\phi_{\rm{MEC}})}}\!,...,e^{j\frac{2\pi}{\lambda}(M-1)\sin{(\phi_{\rm{MEC}})}}]^{H}\!\!,
\end{equation}
respectively. In (25) and (26), $\phi_{\rm{RIS}}$ and $\phi_{\rm{MEC}}$ are AoD or AoA, repectively. The THz channel between the RIS and the $b$th VR user is given by
\begin{equation}
    \tilde{\textbf{g}}_{b}(t) = g_{f,d_{b}}^{\rm{NLoS}}(t)\textbf{a}_{\phi_{b}}^{\rm{NLoS}}(t),
\end{equation}
where $\tilde{\textbf{g}}_{b}(t)=\{{\textbf{g}}_{b}^{\rm{up}}(t), \textbf{g}_{b}(t)\}$, the channel function $g_{f,d_{b}}^{\rm{NLoS}}(t)$ is written as
\begin{equation}
    g_{f,d_{b}}^{\rm{NLoS}}(t) = \frac{c}{4\pi fd_{b}}e^{-\frac{\tau(f)d_{b}}{2}}e^{-j2\pi f\delta_{\text{NLoS},b}(t)},
\end{equation}
$d_{b}$ is the distance between the RIS and the $b$th VR user, and $\textbf{a}_{\phi_{b}}^{\rm{NLoS}}(t)$ is written as 
\begin{equation}
    \textbf{a}_{\phi_{b}}^{\rm{NLoS}}(t) = \frac{1}{\sqrt{N}}[1, e^{j\frac{2\pi}{\lambda}\sin{(\phi_{b})}},...,e^{j\frac{2\pi}{\lambda}(N-1)\sin{(\phi_{b})}}]^{H}.
\end{equation}

\subsection{Quality of Experience Model}
The QoE of the wireless VR video frame streaming can be influenced by several factors, including video quality, VR interaction latency, and smoothness of the VR video frame \cite{XR}. The success of the uplink transmission will further affect the prediction of the viewpoint and the status of LoS or NLoS of the VR user. We use unit-impulse function $\hat{\delta}_k(t)$ to denote the success of the viewpoint prediction, which is expressed as
\begin{equation}
    \hat{\delta}_k(t) = \begin{cases}  {1}, &\text{if } \hat{V}_t^{k} = {V}_t^{k}; \\
    0, & \text{otherwise}.
    \end{cases}
\end{equation}
where $\hat{V}_t^{k} = (\hat{X}_t^{k}, \hat{Y}_t^{k}, \hat{Z}_t^{k})$ and ${V}_t^{k} = ({X}_t^{k}, {Y}_t^{k}, {Z}_t^{k})$ are the predicted and actual viewpoint of the $k$th VR user at the $t$th time slot, respectively. In (30), if $\hat{V}_t^{k} = {V}_t^{k}$, $\hat{\delta}_k(t) = 1$, otherwise, $\hat{\delta}_k(t) = 0$. According to \cite{QoE1} and \cite{QoE2}, the QoE of the $k$th VR user at the $t$th time slot is denoted as
\begin{equation}
    \text{QoE}_{k}(t) =  \hat{\delta}_k(t)(q(R_{k}(t)) - |q(R_{k}(t))- q(R_{k}(t-1))|),
\end{equation}
where $q(R_{k}(t))$ is the VR video transmission quality metrics. Here, due to \cite{QoE2}, $q(R_{k}(t))$ is presented as
\begin{equation}
    q(R_{k}(t)) = \log\left(\frac{R_k^{\text{down}}(t)}{R_{\text{th}}^{\text{down}}}\right),
\end{equation}
where $R_{\text{th}}^{\text{down}}$ is the downlink transmission threshold, and $|q(R_{k}(t))- q(R_{k}(t-1))|$ is the transmission quality variation, which indicates the magnitude of the changes in the transmission quality from the $(t-1)$th time slot to the $t$th time slot. Note that the QoE model in (31) guarantees the seamless, continuous, smoothness and uninterrupted experience of the VR user.

\subsection{Optimization Problem}
To ensure that the requested FoV is rendered and transmitted within the VR interaction latency, we aim to maximize the long-term QoE of the RIS-aided THz transmission system by optimizing the phase shift of the RIS reflecting element under VR interaction latency constraint. At the $t$th time slot, the VR interaction latency $T_{\text{VR}}$ consists of $T_{\text{uplink}}$, $T_{\rm{render}}$, and $T_{\rm{downlink}}$ \cite{Hu1}, which is written as
\begin{equation}
    T_{\text{VR}}(t) = T_{\text{uplink}}(t) + T_{\text{render}}(t) +T_{\text{downlink}}(t),
\end{equation}
where $T_{\text{uplink}}(t)$ is the uplink transmission latency, and $T_{\rm{render}}(t)$ is the MEC rendering latency. For the centralized online GRU, the size of the uplink data is small, and the uplink transmission latency is negligible. It is important to know that the size of FoV does not change for different viewpoints, thus, the rendering latency remains the same. Therefore, the VR interaction constraint condition can be converted to downlink transmission latency constraint. The proposed THz VR system aims at maximizing the long-term total QoE under the downlink transmission latency constraint in continuous time slots with respect to the policy $\pi$ that maps the current state information $S_t$ to the probabilities of selecting possible actions in $A_t$. We formulate the optimization as
\begin{align}
    \max_{\pi(A_t|S_t)}&\sum\limits_{i=t}^{\infty}\sum\limits_{k=1}^{K}\gamma^{i-t}\text{QoE}_{k}(i),\\
    s.t. ~& T_{\text{downlink}}^{k}(i) \leq T_{\text{th}}^{\text{downlink}},
\end{align}
where $\gamma\in [0,1)$ is the discount factor which determines the weight of the future QoE, and $\gamma=0$ means that the agent only considers the immediate reward. In (35), $T_{\rm{downlink}}^{k}(t)$ is the downlink transmission latency of the $k$th VR user at the $t$th time slot, and $T_{\text{th}}^{\text{downlink}}$ is the downlink transmission latency constraint. Note that (35) guarantees the VR interaction latency in each time slot under the VR interaction latency constraint.

Due to the fact that the mobility of the VR user is Markovian in continuous time slots, the dynamics of the THz VR system is a Partially Observable Markov Decision Process (POMDP) problem, which is generally intractable. Here, the partial observation refers to that the MEC server can only know the previous viewpoints and locations of the VR users in the environment, while it is unable to know all the information of the environment, including, but not limited to, the channel conditions, and the viewpoint in the current time slot. Meanwhile, the selected policy also needs to satisfy the VR interaction threshold constraint. Thus, the problem in (34) is a constrained MDP (C-MDP) problem that can be transformed into the following form
\begin{equation}
\begin{aligned}
    \!\!\min_{\omega\geq 0, \mu\geq 0} \!\!\!\!\max_{\pi} \!\!\sum\limits_{i=t}^{\infty}\sum\limits_{k=1}^{K}&\gamma^{i-t}\text{QoE}_{k}(i) \!-\! \omega(T_{\text{th}}^{\text{downlink}} \!\!-\!\! T_{\text{downlink}}^{k}(i)),
\end{aligned}
\end{equation}
where $\omega$ is the Lagrangian multiplier, and $\pi$ is the policy. Due to the fact that the number of combinations of the phase shift increases exponentially with the number of phase shift levels of the RIS, the problem in (36) is the generalization of large dimension C-MDP. To address this issue, we deploy constrained deep reinforcement learning (C-DRL) to solve this problem in (36) in Section IV.

\begin{figure}[!h]
    \centering
    \includegraphics[width=3.5 in]{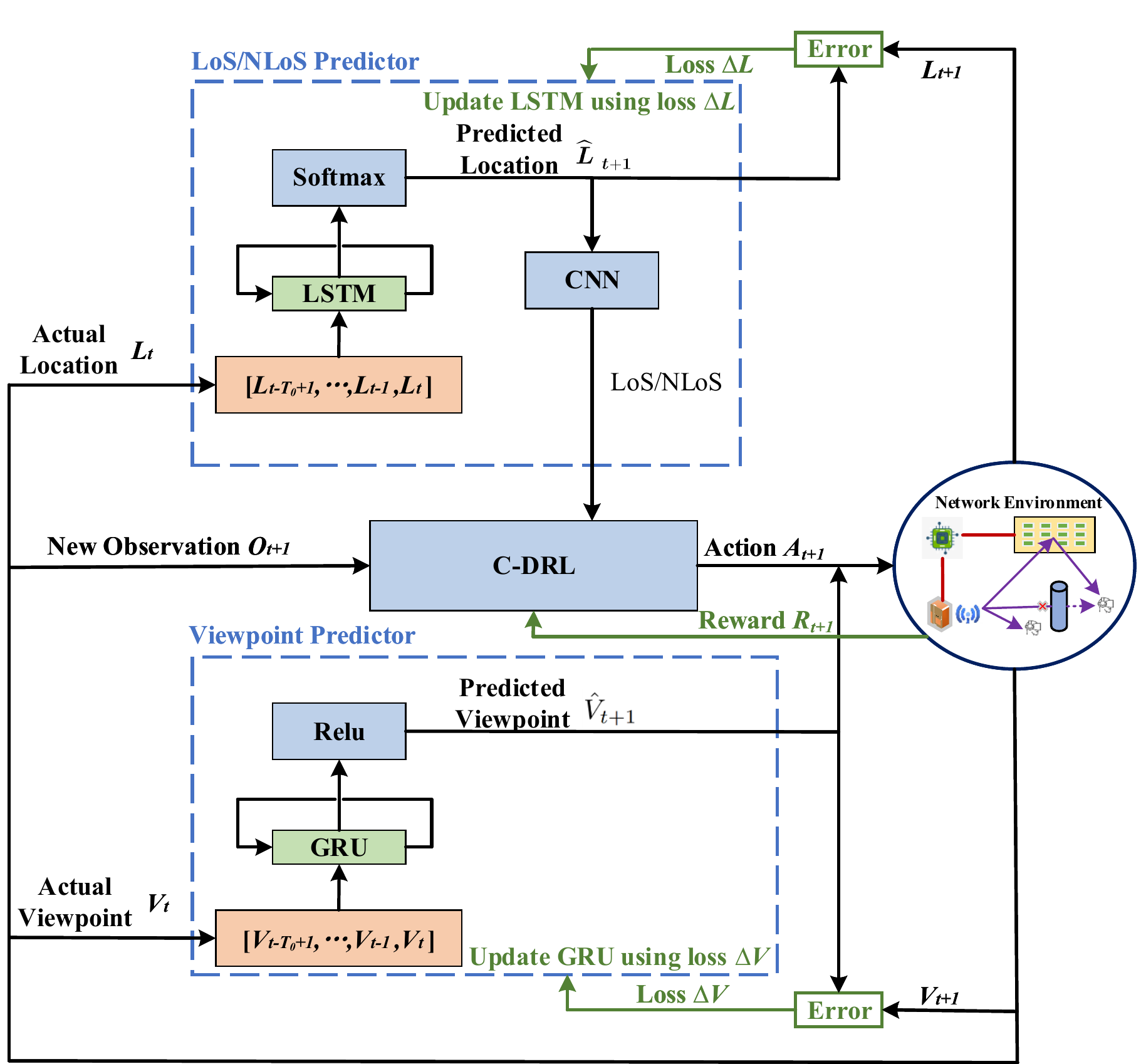}
    \caption{Learning strategy for MEC-enabled and RIS-assisted THz VR networks.}
    \label{basic_modules}
\end{figure}

\section{Learning Algorithms for THz VR System}
The deep neural network is one of the most popular non-linear approximation functions, and C-DRL can effectively solve C-MDP problem \cite{CDRL1}. To solve the optimization problem in (36), we propose a novel learning architecture based on online GRU, online LSTM, CNN, and C-DRL, as shown in Fig. 5. In particular, the online GRU and online LSTM are integrated with CNN to predict the viewpoint preference and LoS or NLoS status of each VR user in continuous time slots, respectively. Using this information as inputs, the C-DRL is deployed to select an optimal reflection coefficient matrix for THz downlink transmission.

\subsection{Viewpoint Prediction}
We use the centralized online GRU and distributed FedAvg to predict the viewpoints of the VR users over time. The input of the learning model is the actual viewpoints of the previous time slots, and the output is the predicted viewpoint of the VR user for the next time slot. For the centralized online GRU, the VR user directly transmits its actual viewpoint to the MEC through uplink transmission at each time slot, and then the viewpoint is predicted based on the current and previous time slots. The centralized online GRU for viewpoint prediction has already been introduced in Section IV of \cite{XNL} in detail.

For the federated learning among distributed VR devices, each VR user predicts the viewpoint in its VR device in continuous time slots based on the online GRU algorithm. At each time slot, the updated learning model of each VR user is delivered to the MEC for model aggregation via uplink transmission, and the model aggregation at the MEC at the $t$th time slot can be denoted as
\begin{equation}
    \bar{{\bm{\theta}}}_{t}^{\text{GRU}} = \sum_{k=1}^{K}p_k{\bm{\theta}}_{k,t}^{\text{GRU}},~\text{and}~p_k = \frac{n_k}{n},
\end{equation}
where $p_k$ is the percentage of the number of data samples of the $k$th device in the total number of data samples, $n_k$ is the number of data samples of the $k$th VR user, and $n$ is the total number of data samples, which is calculated as $n = \sum_{k=1}^{K}n_k$. Then, the aggregated model is transmitted to each VR user through downlink transmission to predict the viewpoint, and the predicted viewpoint is delivered to the MEC for VR video frame rendering. Due to the fact that the size of the learning model is much larger than that of the viewpoint, the VR interaction latency may be increased.

\subsection{LoS and NLoS Prediction}
When VR users move in the indoor scenario following the VRMM mobility model, they may be blocked by the VR users with higher height or the obstacle. To predict the LoS or NLoS status of each VR user in continuous time slots, we first employ an RNN model based on LSTM to predict the position of the VR user \cite{LSTM_mobility}. Then, we map the indoor scenario into a 2D image, label the positions of the MEC, the VR users, and the obstacles with different colors, and deploy the CNN to predict the LoS or NLoS status of each VR user.

\subsubsection{Long-short Term Memory}
To capture the dynamics in mobility of the VR user for the $(t+1)$th time slot, both of the most recent observation ${O}_t = \{{O}_t^1,{O}_t^2,...,{O}_t^{K}\}$ and the previous observations ${H}_t = \{{O}_{t-T_o+1},...,{O}_{t-2},{O}_{t-1}\}$ given a memory window $T_o$ are required, where ${O}_t^k = L_t^k = [X_t^k, Y_t^k, H_t^k]$ is the actual location of the $k$th VR user at the $t$th time slot. In the VRMM, we assume that there are four moving directions, which are up, down, left and right, and can be denoted as $D_{\text{u}}$, $D_{\text{d}}$, $D_{\text{l}}$, and $D_{\text{r}}$, respectively. To detect the the moving direction of the VR user over time, an RNN model with parameters ${\bm{\theta}}^{\text{RNN}}$, and a LSTM architecture in particular, is leveraged, where ${\bm{\theta}}^{\text{RNN}}$ consists of both the LSTM internal parameters and weights of each layer.

The online LSTM layer has multiple standard LSTM units and receives current and previous observations at each time slot via the two-ray uplink transmission, and is connected to an output layer, which consists of a Softmax non-linearity activation function with four output values. The four output values represent the predicted probabilities $\mathcal{P}\{D_t = \{D_{\text{u}}, D_{\text{d}}, D_{\text{l}}, D_{\text{r}}\}|[{O}_t^1,{O}_t^2,...,{O}_t^{K}], {\bm{\theta}}^{\text{RNN}}\}$ of the moving directions for the $t$th time slot with historical observations $[{O}_t^1,{O}_t^2,...,{O}_t^{K}]$. 

To update the model parameter ${\bm{\theta}}^{\text{RNN}}$, a standard Stochastic Gradient Descent (SGD) \cite{SGD1} via BackPropagation Through Time (BPTT) \cite{BPTT} is used. At the $(t+1)$th time slot, the parameters ${\bm{\theta}}^{\text{RNN}}$ are updated as
\begin{equation}
    {\bm{\theta}}_{t+1}^{\text{RNN}} = {\bm{\theta}}_{t}^{\text{RNN}} - {\lambda}^{\text{RNN}}\nabla {\mathcal{L}}^{\text{RNN}}({\bm{\theta}}_{t}^{\text{RNN}}),
\end{equation}
where ${\lambda}^{\text{RNN}}\in(0,1]$ is the learning rate, $\nabla {\mathcal{L}}^{\text{RNN}}({\bm{\theta}}_{t}^{\text{RNN}})$ is the gradient of the loss function ${\mathcal{L}}^{\text{RNN}}({\bm{\theta}}_{t}^{\text{RNN}})$ to train the RNN predictor. Here, ${\mathcal{L}}^{\text{RNN}}({\bm{\theta}}_{t}^{\text{RNN}})$ is obtained by averaging the cross-entropy loss as 
\begin{equation}
    {\mathcal{L}}_{t}^{\text{RNN}}({\bm{\theta}}^{\text{RNN}}) \!=\! -\!\!\!\!\!\!\!\!\sum_{t^{'} = t-T_b+1}^{t}\!\!\!\!\!\!\!\!\!\log\left(\!\mathcal{P}\{D_{t^{'}} = \hat{D} |O_{t^{'}-T_{0}}^{t^{'}},{\bm{\theta}}^{\text{RNN}}\}\!\right),
\end{equation}
where
\begin{equation}
    \hat{D} \!\!= \!\!\{D_{\text{u}}, D_{\text{d}}, D_{\text{l}}, D_{\text{r}}\},~ {O}_{t^{'}-T_{0}}^{t^{'}} \!\!\!\!=\! [{O}_{t^{'}-T_{0}+1},...,{O}_{t^{'}-1},{O}_{t^{'}}],
\end{equation}
and $T_b$ is the randomly selected mini-batch size.

By predicting the moving direction of each VR user, the MEC is able to know the position of the VR user in the next time slot in advance. 

\subsubsection{Convolutional Neural Network} Based on the predicted positions of VR users, the MEC, obstacles, and the VR users with different heights are labeled with different colors, and mapped into a 2D image to be the input of the CNN. The CNN predicts the LoS/NLoS status of each VR user based on the input 2D image. 

\begin{figure}[!h]
    \centering
    \includegraphics[width=3.5 in]{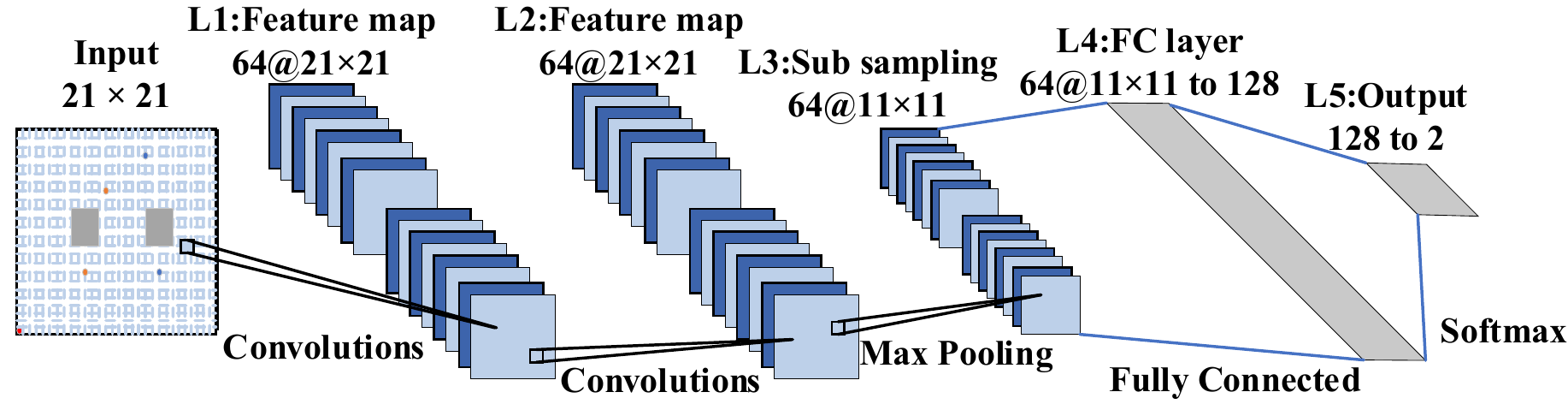}
    \caption{Proposed CNN to classify the LoS or NLoS status of VR users.}
    \label{basic_modules}
\end{figure}

The CNN is a multi-layer network evolved from the traditional neural network. The CNN mainly includes the input layer, convolution layer, pooling layer, fully-connected layer, and output layer. It is used for feature extraction and mapping through fast training, and possesses high classification and prediction accuracy. We assume that the proposed CNN model consists of one data input layer, $N_c$ convolution layers, $N_p$ pooling layers, $N_f$ fully-connected layers, and one output layer. Meanwhile, it is assumed that the size of the input image is $N_0\times N_0$. The detailed description of each layer in the CNN is introduced as follows:

(a) Data Input Layer: The MEC, obstacles, VR user with higher height, and VR user with lower height are denoted by different colors in the 2D image. The preprocessed images are used as the input data of the convolution network, and the initial feature extraction is obtained via principal component analysis (PCA) \cite{PCA}, which is usually used for feature extraction. Meanwhile, the images are projected into the characteristic subspace of $N_0\times N_0\times 3$, where $3$ presents the color of the image is in RGB mode. Therefore, the learning efficiency and computational complexity of the CNN can be decreased by reducing the image dimensionality.

(b) Convolution Layer: The characteristics of the input image is extracted by a randomly initialized filter. It is possible that the input image has various characteristics, thus, multiple filters are used to extract all features in the original image. Zero padding is also used for each convolution layer to keep the size of the features extracted from the input image as $N_0\times N_0$.

(c) Pooling Layer: It plays an important role in sub-sampling via using the features extracted from the convolution layers. The time complexity can be decreased in the next convolution layer or fully-connected layer by reducing the number of operations in sub-sampling. The Max-pooling method is usually deployed to extract the largest value in the sliding window for the sub-sampling among all the methods used in the pooling layer.

(d) Fully-connected Layer: The features extracted by the convolution and pooling layer are inserted into the neural network. A softmax layer that is often used for the classifications of multiple classes is employed at the end of the fully-connected layer. Meanwhile, the classification result corresponds to a probability that the sum of the probabilities of all classes is equal to 1, and the class with the highest probability is the estimated label for the corresponding input image.

For example, the proposed CNN to classify the LoS or NLoS statuses of VR users is shown in Fig. 6, which is also used for the simulations in Section V. In Fig. 6, a $21\times 21$ image goes through two convolution layers with 64 filters, and one max-pooling layer. The kernel size and pooling size are $2\times 2$. Then, the extracted features pass through a fully-connected layer with 128 filters. The activation function in the convolution layer and fully-connected layer is the ReLu function. After passing through the fully-connected layer, the softmax layer is used to determine the LoS or NLoS status corresponding to the input image. Note that Adam Optimizer is also employed while training the proposed CNN model \cite{adam}.

\subsection{Downlink RIS Configuration}
The main purpose of Reinforcement Learning (RL) is to select proper reflection coefficient matrix $\boldsymbol{\Theta}$ given in (14) of the RIS for THz downlink transmission for the VR users in the NLoS area. While for the uplink transmission at the $(t+1)$th time slot, it directly uses the selected $\boldsymbol{\Theta}$ at the $t$th time slot. This is because the downlink transmission requires a high data rate for the FoV with high resolution, whereas the uplink transmission only transmits the actual position and viewpoint or the learning model of the VR user (e.g., the size of the uplink data is much smaller than that of the FoV). Through a series of action strategies, the MEC is able to transmit the selected $\boldsymbol{\Theta}$ to the RIS via wired connection, interact with the environment, and obtain rewards based on its actions, which can help improve the action strategy. With enough number of iterations, the MEC is able to learn the optimal policy that maximizes the long-term reward.

We use $S\in\mathcal{S}$, $A\in\mathcal{A}$, and $R\in\mathcal{R}_{e}$ to denote the state, action and reward from their corresponding sets, respectively. The purpose of the RL algorithm is to find an optimal policy $\pi$ to maximize the long-term reward for $A = \pi(S)$. The optimization function can be formulated as $<S, A, R>$, and the detailed descriptions of the state, action, and reward of the optimization problem in (34) are introduced as follows.
\begin{itemize}
    \item State: At the $t$th time slot, the network state is denoted as 
    \begin{align}
        S_{t} &= (\mathcal{L}_{t}, \mathcal{I}_{t}, \widehat{\text{QoE}}_{t-1})\in \mathcal{S},  \\ \nonumber
        \text{with}~
        \mathcal{L}_{t} &= \{ {L}_{t}^{1}, {L}_{t}^{2},...,  {L}_{t}^{K^{\text{VR}}} \},\\ \nonumber
        \mathcal{I}_{t} &= \{{I}_{t}^{1}, {I}_{t}^{2},...,  {I}_{t}^{K^{\text{VR}}}\} \\ \nonumber
        \widehat{\text{QoE}}_{t-1} &= \{{\text{QoE}}_{t-1}^{1}, {\text{QoE}}_{t-1}^{2},...,  {\text{QoE}}_{t-1}^{K^{\text{VR}}}\},
    \end{align}
    where $\mathcal{L}_{t}$ is the set containing the positions of all VR users at the $t$th time slot, where ${L}_{t}^{k} = [X_t^k, Y_t^k, H_t^k]$, ${I}_{t}^{k} = \{1, 0\}$ is the predicted LoS or NLoS status of the $k$th VR user for the $t$th time slot, where 1 represents LoS, 0 represents NLoS, and ${\text{QoE}}_{t-1}^{k}$ is the QoE value calculated by (31) of the $k$th VR user for the $(t-1)$th time slot.
    
    \item Action: The action space is written as
    \begin{align}
        {A}_{t} &= \{\widetilde{\boldsymbol{\Theta}}_{t}\}\in\mathcal{A}, \\ \nonumber
        \text{with}~\widetilde{\boldsymbol{\Theta}}_{t} &= \{\boldsymbol{\Theta}_{t}^{1}, \boldsymbol{\Theta}_{t}^{2},...,\boldsymbol{\Theta}_{t}^{\hat{L}^{N}}\},
    \end{align}
    where $\widetilde{\boldsymbol{\Theta}}_{t}$ is the set that includes all the possible reflection coefficient matrix given the number of reflection elements $N$ of the RIS and the number of phase shift levels $\hat{L}$, with $\boldsymbol{\Theta}_{t}^{i}$ given in (14).
    
    \item Reward: The immediate reward $R_t$ is designed as
    \begin{equation}
        {R}_{t}(S_t,A_t) = \sum\limits_{k=1}^{K_{\text{VR}}} \text{QoE}_{t}^{k}.
    \end{equation}
\end{itemize}

The performance of the selected action is determined by the position and LoS/NLoS status of the VR user, which can further influence the long-term QoE of the THz VR system. Therefore, we use the observed position, the LoS/NLoS status, and the QoE of the VR user as observation, and use the QoE as a reward. According to the observed environmental state $S_t$ at the $t$th time slot, the MEC selects specific action $A_t$ from the set $\mathcal{A}$ and obtains reward $R_t$. Then, the discounted accumulation of the long-term reward is denoted as
\begin{equation}
    {Q}(S,\pi) = \sum_{i = t}^{\infty}(\gamma)^{i - t}{R}_{i}(S_i,A_i),
\end{equation}
where $\gamma\in[0,1)$ is the discount factor.

When the number of reflection elements and phase shift levels is small, the RL algorithm can efficiently obtain the optimal policy. However, when a large number of reflection elements and phase shift levels exist, e.g. $10^{50}$ ($\hat{L} = 10$, $N = 50$), the state and action spaces will be increased proportionally, which will not only occupy plenty of computation memory of the MEC, but also inevitably result in massive computation latency and degraded performance of the RL algorithm. To address this issue, deep learning is introduced to RL, namely, deep reinforcement learning (DRL), through interaction with the environment, DRL can directly control the behavior of the MEC, and solve complex decision-making problems. Meanwhile, the policy should not violate the VR interaction latency threshold. Thus, we use a constrained deep Q network (C-DQN) to solve the optimization problem, which indirectly optimizes the policy by optimizing the value function while satisfying the downlink latency constraint.

\begin{figure}[!h]
    \centering
    \includegraphics[width=3.5 in]{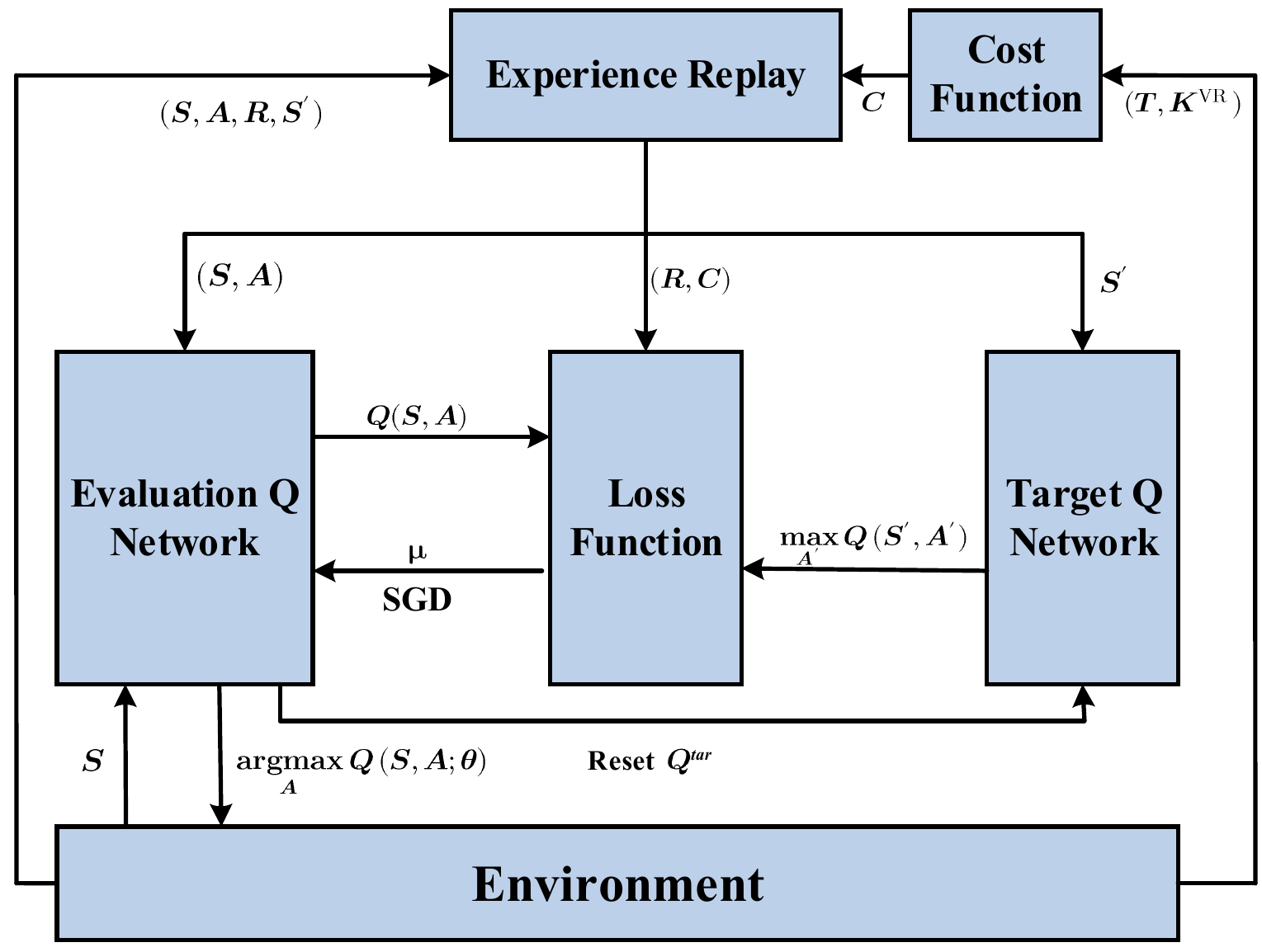}
    \caption{The C-DRL diagram of the THz transmission scheme.}
    \label{basic_modules}
\end{figure}

As shown in Fig. 7, C-DQN is a value-based DRL algorithm, it combines a neural network with Q-learning and optimizes the state-action value function through a deep neural network (DNN). The C-DQN uses a neural network to store state and action information $Q^{\pi}(S, A)$. It also applies the experience replay to train the learning model, and experiences in the experience replay are partially selected to learn to improve the learning efficiency of the neural network and break the correlation among the training samples. In addition, the distribution of the training samples can be smoothed via averaging the selected samples, which can further avoid the training divergence.

The objective of C-DQN is to find the optimal policy $\pi^{\star}$, and obtain optimal state-action value $Q^{\star}(S, A)$, which is expressed as
\begin{equation}
    \pi^{\star}(S) = \arg\max_{A} Q^{\star}(S, A).
\end{equation}
In the proposed optimization problem, C-DQN should satisfy the VR downlink transmission latency constraint. Thus, the state-action value is calculated as
\begin{equation}
    Q(S, A) = R + \gamma_{\text{CDQN}}\max_{A^{'}}Q(S^{'}, A^{'}) - \mu C^{\text{down}},
\end{equation}
where $S^{'}$ is the next state, $A^{'}$ is the next action, and $\gamma_{\text{CDQN}}$ is the discount factor, which determines the balance between the current state-action value and future state-action value. In (46), $C^{\text{down}}$ is the downlink transmission cost due to the constraint in (35) at each time slot, which is calculated as
\begin{equation}
    C^{\text{down}} = T_{\text{th}}^{\text{downlink}} - \frac{\sum_{k=1}^{K^{\text{VR}}}T_{\text{downlink}}^{k}}{K^{\text{VR}}},
\end{equation}
where $K^{\text{VR}}$ is the number of the VR users, $T_{\rm{downlink}}^{k}(t)$ is the downlink transmission latency of the $k$th VR user at the $t$th time slot, and $T_{\text{th}}^{\text{downlink}}$ is the downlink transmission latency constraint. According to (46), the Q evaluation network in C-DQN is used to estimate $Q(S, A)$. Note that the target Q network does not change in each time slot and is updated after several time slots. To update the evaluation state-action value, Bellman Equation is applied, which is denoted as
\begin{equation}
    Q^{e}(S, A) = (1-\alpha_{\text{CDQN}})Q^{e}(S, A) + \alpha_{\text{CDQN}} Q^{tar}(S, A),
\end{equation}
where $Q^{e}$ and $Q^{tar}$ are the output of Q evaluation and target network, respectively. In (48), $\alpha_{\text{CDQN}}$ is the learning rate. The loss function is calculated as $(Q^{tar} - Q^{e})$, which is used to update the weights of the Q evaluation network. 
We can obtain the optimal policy and Q value when the C-DRL converges. Our detailed C-DRL algorithm is presented in Algorithm 1.

\begin{algorithm}[t]
\begin{algorithmic}[1]
\caption{C-DRL to select the optimal phase shifts of the RIS in THz transmission}
\STATE Initialize replay memory $G$,  discount factor $\gamma_{\text{CDQN}}\in[0,1)$, and learning rate $\alpha_{\text{CDQN}}\in(0,1]$.
\STATE Initialize state-action value function ${Q}(S,A)$, the parameters of evaluation Q network and target Q network.
\FOR{Iteration = 1,...,$I$}
    \STATE Input the network state $S$.
    \FOR{t = 1,...,T}
        \STATE Use $\epsilon$-greedy algorithm to select a random action $A_t$ from the action space $\mathcal{A}$.
        \STATE Otherwise, select $A_t = \max\limits_{A\in \mathcal{A}}{Q}(S_t,A)$.
        \STATE The MEC performs downlink transmission according to the selected action $A_t$.
        \STATE The MEC observes reward $R_t$, new state $S_{t+1}$ and calculates the cost according to (47).
        \STATE Store transition $(S_{t},A_{t},R_{t},C_{t}^{\text{down}}, S_{t+1})$ in replay memory $G$.
        \STATE Sample random minibatch of transitions $(S_j,A_j,R_j,C_{j}^{\text{down}}, S_{j+1})$ from replay memory $G$.
        \IF{$j+1$ is terminal}
            \STATE $y_j^{target} = R_j$.
        \ELSE
            \STATE $y_j^{target} = R_{j+1} + \gamma\max\limits_{A}{Q}(S_{j+1},A)$.
        \ENDIF
        \STATE Update evaluation Q  network.
        \STATE Update the Lagrangian multiplier with
        \begin{equation}
            \omega = \omega + \alpha_{\text{CDQN}}\frac{1}{|G|}\sum\limits_{i=1}^{|G|}C_{i}^{\text{down}}, 
        \end{equation}
        where $|G|$ is the size of the replay memory.
        \STATE Update target Q  network periodically. 
    \ENDFOR
\ENDFOR
\end{algorithmic}
\end{algorithm}

\begin{table*}
\centering
\caption{Simulation Parameters of THz VR Network}
\begin{tabular}[c]{c|c|c|c}
\hline
\hline Indoor scenario size & 20~$\mathrm{m}$ $\times$ 20~$\mathrm{m}$ $\times$ 3~$\mathrm{m}$ & Number of MEC& 1 \\
\hline Location of MEC & [0, 0, 3~$\mathrm{m}$] & Number of VR users & 5 \\
\hline Height of VR user & [1.2~$\mathrm{m}$, 1.8~$\mathrm{m}$] & Location of obstacle (X axis) & [4~$\mathrm{m}$, 8~$\mathrm{m}$], [12~$\mathrm{m}$, 16~$\mathrm{m}$]\\
\hline Location of obstacle (Y axis) & [8~$\mathrm{m}$, 12~$\mathrm{m}$] & Height of obstacle (Z axis) & 3~$\mathrm{m}$\\ 
\hline Center Location of RIS & [10~$\mathrm{m}$, 20~$\mathrm{m}$, 3~$\mathrm{m}$] & Speed of Light & $3\times 10^8$ $\mathrm{m/s}$\\
\hline THz center frequency & 300$~\text{GHz}$ & Number of phase shift elements & 20\\
\hline FoV resolution & 4k & MEC execution ability $F_{\text{MEC}}$ & $5~\text{GHz}$\\
\hline Number of cycles processing one bit $f_{\text{MEC}}$ & $1000~ \text{Cycles/bit}$ & Number of antennas of MEC & 30 \\
\hline Downlink transmission latency & 12 $\text{ms}$ & White Gaussian noise $\sigma^{2}$ & $-110~\text{dBm}$ \\
\hline LSTM memory size & 10 & Minibatch size & 64 \\
\hline RNN learning rate & 0.005 & Discount Factor $\gamma$ & 0.9 \\
\hline Number of C-DRL layer & 2 & Number of C-DRL units of each layer & 128  \\
\hline C-DRL Learning rate $\alpha_{\text{CDQN}}$ & 0.05 & Time slots & 300 \\
\hline
\hline
\end{tabular}
\end{table*}

\subsection{Computational Complexity Analysis of Learning Algorithms}
For the computation complexity of the RNN based on the GRU and LSTM architecture, it is computed as $O(\widetilde{m}\widetilde{n}\log \widetilde{n})$, where $\widetilde{m}$ is the number of layers, and $\widetilde{n}$ is the number of units per learning layer. The computation complexity of the CNN is written as $O(\sum_{i=1}^{L_{\text{CNN}}}\hat{m}_{i}^2\hat{n}_{i}^2c_{\text{in}}c_{\text{out}})$,  where $\hat{m}$ is the length of the output feature map of the Convolution kernel, $\hat{n}$ is the length of the Convolution kernel, $c_{\text{in}}$ is the number of the input channels, $c_{\text{out}}$ is the number of the output channels, and $L_{\text{CNN}}$ is the number of CNN layers \cite{CNN_complexity}. The computational complexity of the C-DRL algorithm is given by $O(\bar{m}\bar{n}\log \bar{n})$. Here, $\bar{m}$ is the number of layers, and $\bar{n}$ is the number of units per learning layer \cite{introduction_algorithm}. 

To compare with the C-DRL algorithm, we use an exhaustive algorithm to select the optimal phase shift of the RIS, and the computational complexity of the exhaustive algorithm is denoted as $O(\hat{L}^{N})$ \cite{Nano}, where $\hat{L}$ and $N$ are the number of phase shift levels and the number of reflecting elements of the RIS, respectively.

\begin{figure}[!h]
    \centering
    \includegraphics[width=3.0 in]{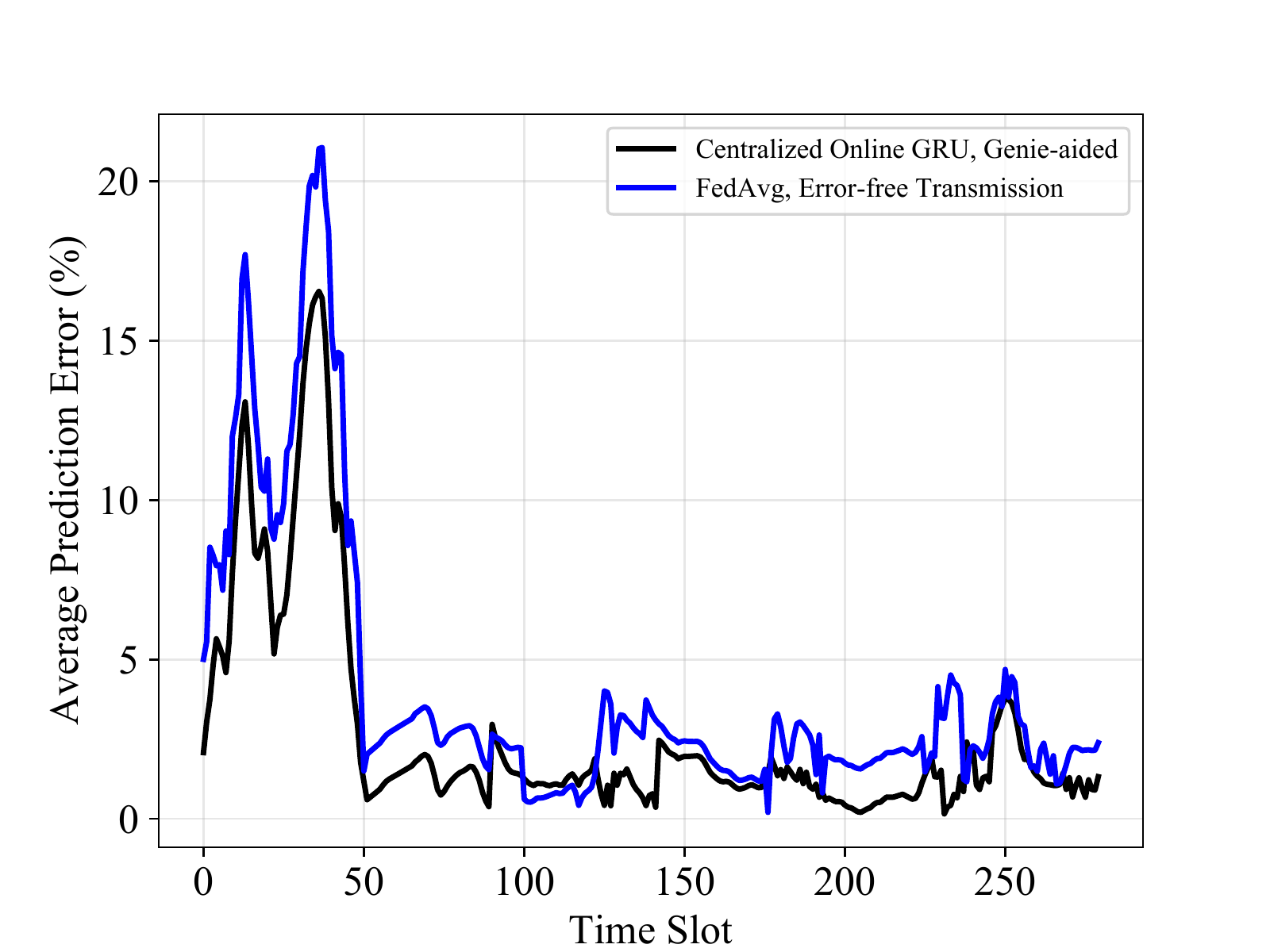}
    \caption{Average prediction error of centralized online GRU and distributed FedAvg algorithms in continuous time slots.}
    \label{basic_modules}
\end{figure}

\section{Simulation Results}
In this section, we examine the effectiveness of our proposed learning architecture in Fig. 5. The simulation parameters are summarized in Table I.

\subsection{Viewpoint Prediction}
The VR dataset obtained from \cite{VRdataset} includes $16$ clips of VR videos with $153$ VR users, and 969 data samples of the motion in three dimensions, pitch, yaw, and roll, namely, $X$, $Y$ and $Z$ viewing angles. The viewpoint ranges of $X$, $Y$ and $Z$ angles are (-$50^{\circ}$, $50^{\circ}$), (-$150^{\circ}$, $150^{\circ}$) and (-$50^{\circ}$, $50^{\circ}$), respectively. According to \cite{Bao2}, the motion of the VR user has strong short-term auto-correlations in all three dimensions. Due to the fact that auto-correlations are much stronger than the correlation between these three dimensions, the angles in each direction can be trained independently and separately. In addition, the range of $Y$ angle distribution is much larger than that of $X$ and $Z$. Therefore, for simplicity, we use online GRU to predict the $Y$ angle of VR users in this section, however, our algorithms can also be used for the prediction of $X$ and $Z$ angles. In the simulation, we use the viewpoint samples of the historical ten time slots (1 second) to predict the viewpoint of the next time slot (0.1 seconds).

Fig. 8 plots the average prediction error of the centralized online GRU and distributed FedAvg algorithms in continuous time slots. In the Genie-aided scheme, the learning algorithms are trained with the known correct actual viewpoint of each VR user by the MEC at each time slot. In the Error-free Transmission scheme, there are no transmission errors in the uplink and downlink transmission. We can see that the performance of the centralized online GRU is better than that of the FedAvg. This is because the FedAvg depends on the local data of each VR user while minimizing the loss function, and this biases the learning model to be fit for the specific VR user \cite{Agnostic}, whereas the centralized online GRU can learn from global data of all VR users, so that the learning model can be appropriated for all VR users. It is also noted that at the beginning, there are large fluctuations in the performance of the learning algorithms. This is because the parameters in the learning algorithms should be modified to capture the viewpoint preference of the VR user. In addition, more simulation results of viewpoint prediction have already been described in Section V of \cite{XNL} in detail.

\subsection{LoS and NLoS Prediction}
During the LoS and NLoS prediction, the algorithm that integrates online LSTM and CNN is deployed to predict the mobility of VR users and judge the LoS/NLoS status of the VR user in continuous time slots. We first use Python to simulate the mobility of VR users in the indoor scenario, and label the moving direction based on the corresponding mobility data. Then, we use the created mobility dataset to train parameters of the LSTM, which can be further used for the online mobility prediction. Fig. 9 (a) plots the loss of the LSTM of each epoch. It is seen that the LSTM converges after 60 epochs. Fig. 9 (b) plots the prediction error of the LSTM via different number of moving periods. It is noted that when we use the mobility data of the previous 10 time slots to predict the mobility direction for the next time slot, we can obtain the minimum prediction error \cite{XNL}.

\begin{figure}[!t]
	\centering
	\subfloat[]{\includegraphics[width=1.7in]{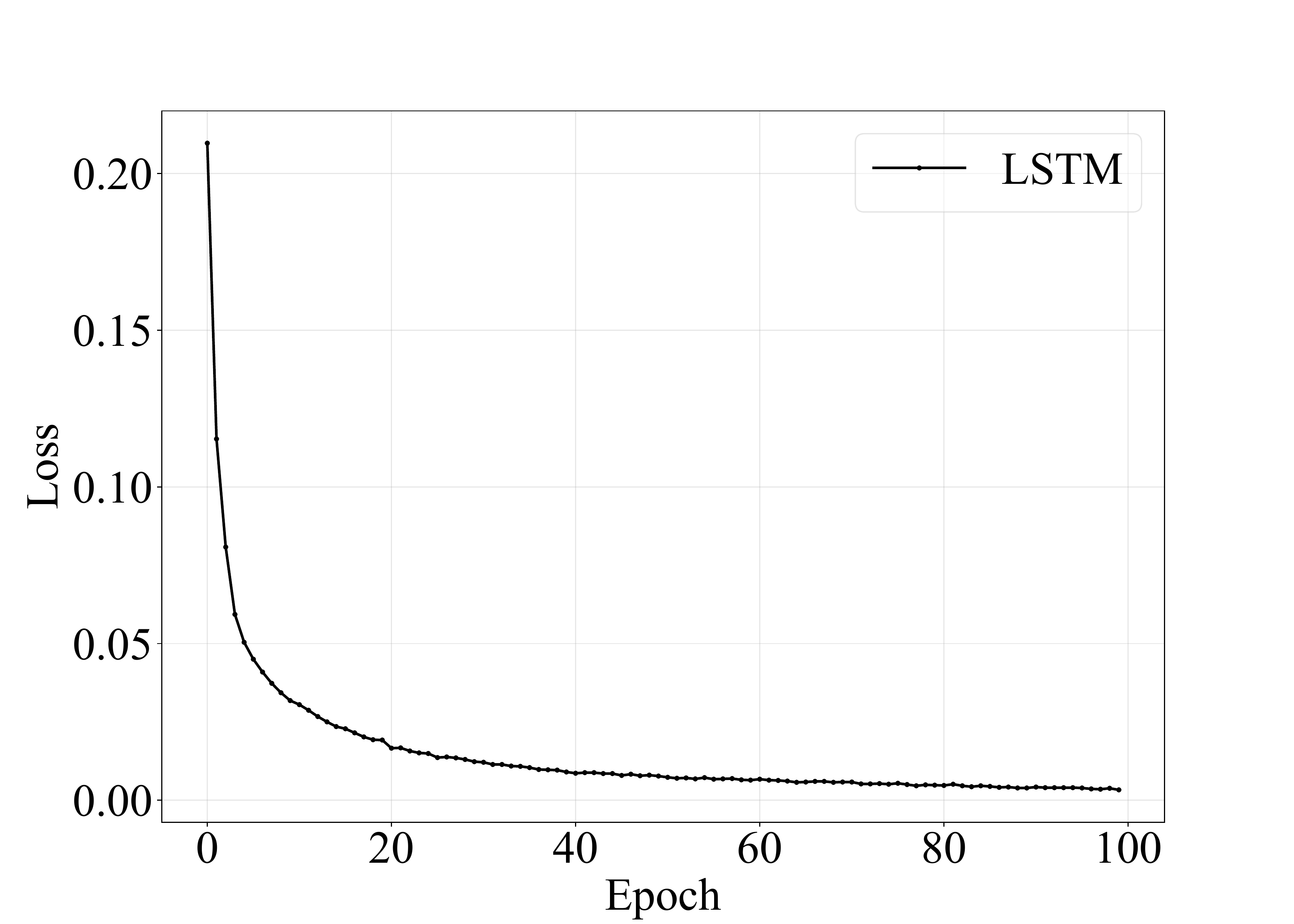}}\label{fig_first_case}
	\hfil
	\subfloat[]{\includegraphics[width=1.7in]{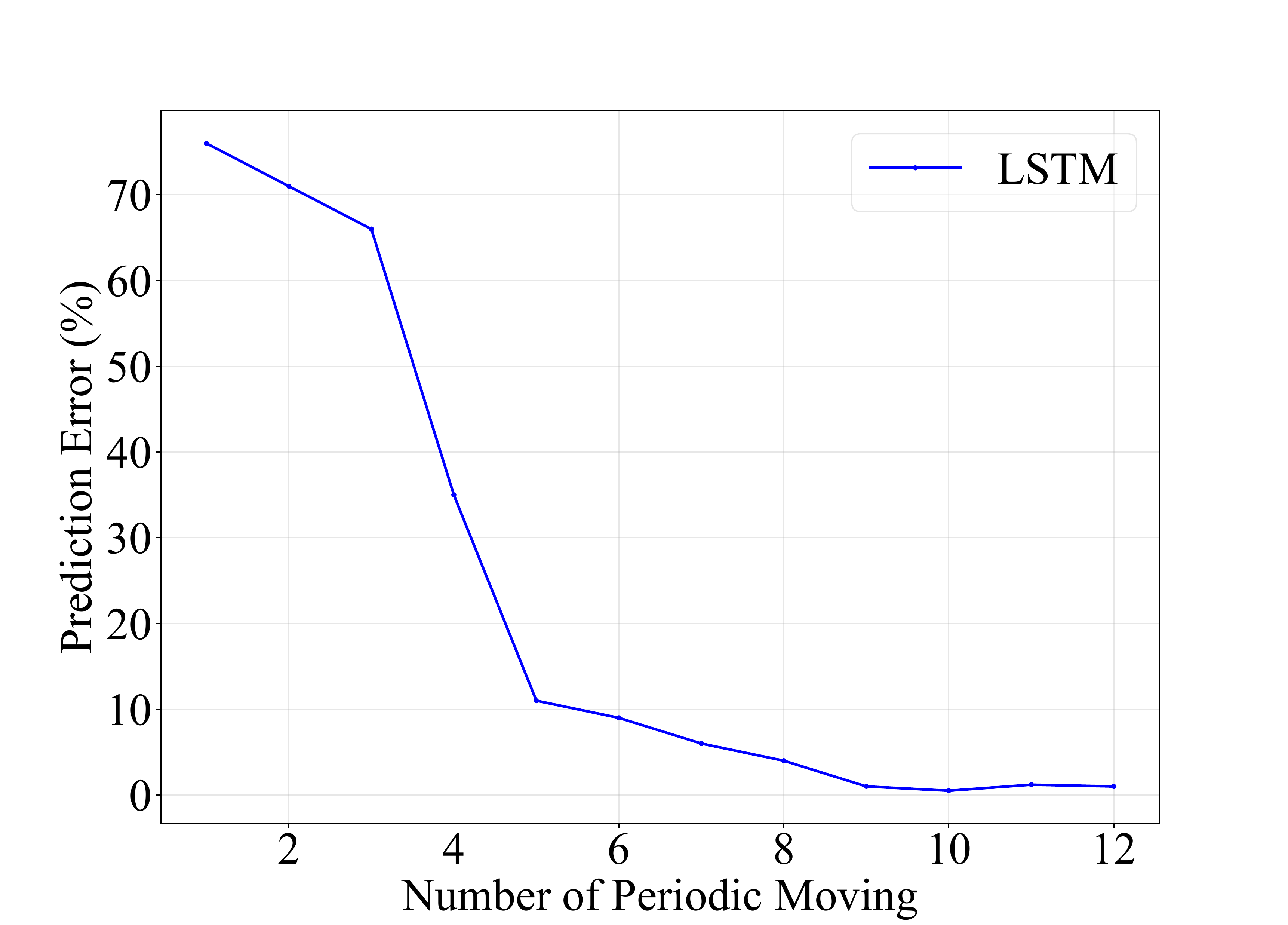}}\label{fig_second_case}
	\caption{(a) Loss of LSTM for VR user mobility prediction of each epoch. (b) Prediction error of LSTM algorithm via different number of moving periods.}
	\label{basic_modules}
\end{figure}

\begin{figure}[!t]
	\centering
	\subfloat[]{\includegraphics[width=1.7in]{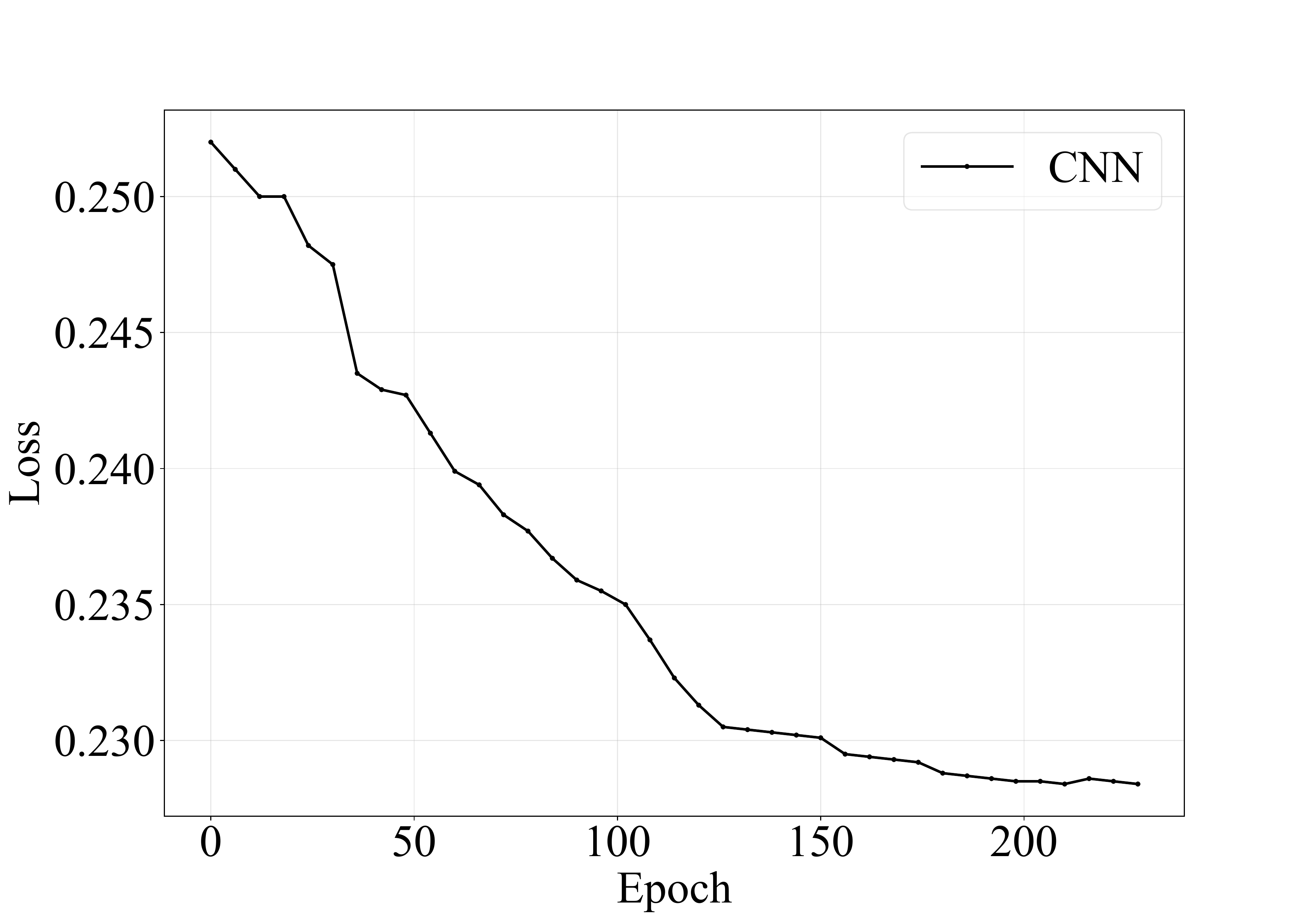}}\label{fig_first_case}
	\hfil
	\subfloat[]{\includegraphics[width=1.7in]{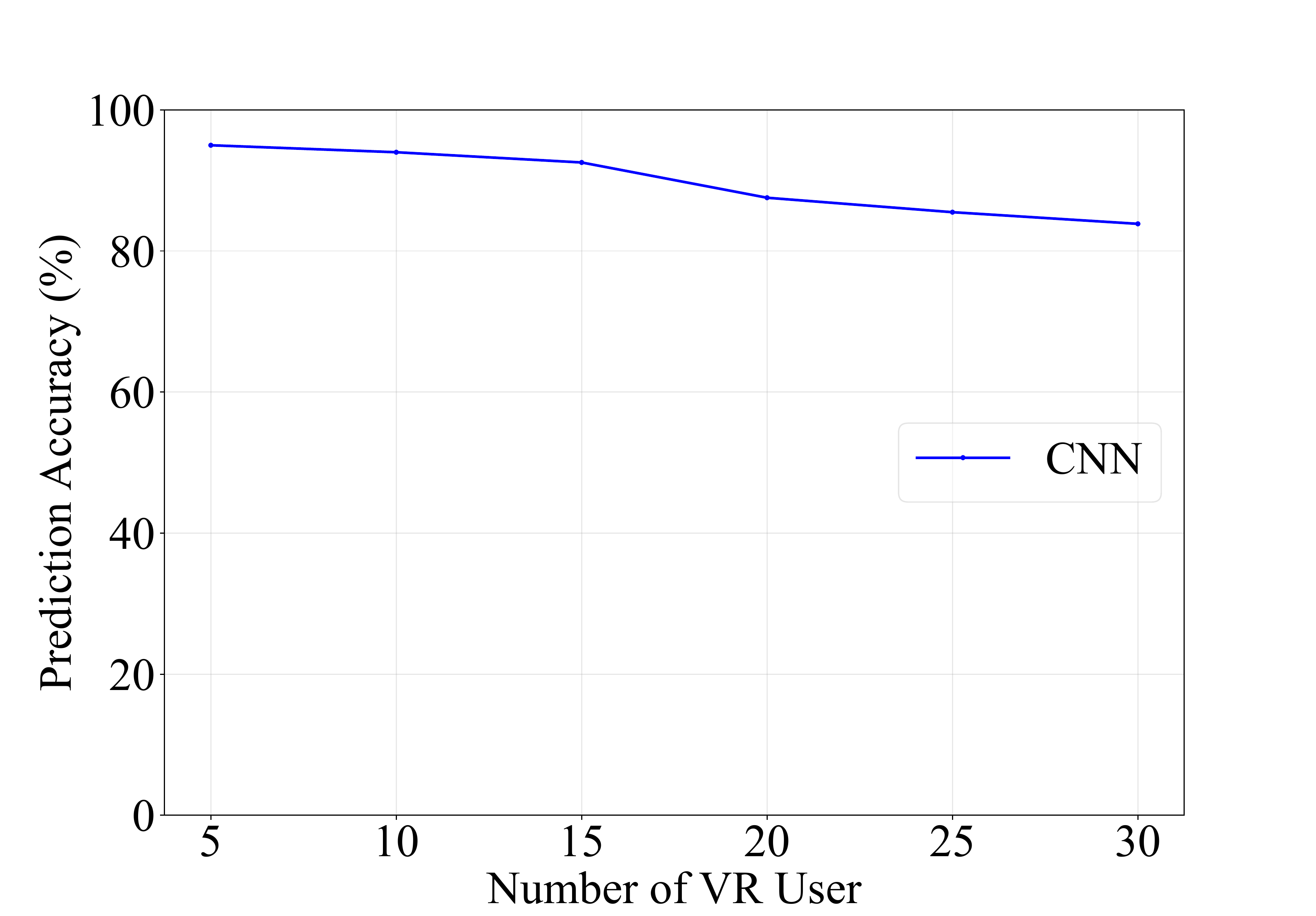}}\label{fig_second_case}
	\caption{(a) Loss of CNN for LoS or NLoS prediction of each VR user of each epoch. (b) Prediction accuracy of CNN algorithm via different number of VR users.}
	\label{basic_modules}
\end{figure}

Fig. 10 (a) plots the loss of CNN for LoS or NLoS prediction of each VR user of each epoch. It is obtained that the CNN converges after 150 epochs. Fig. 10 (b) plots the prediction accuracy of CNN via different number of VR users. We observe that the prediction accuracy is about $95\%$ when the number of VR users is smaller than 15, but decreases with the increasing number of VR users. This can be explained by the fact that when more VR users exist in the indoor scenario, the features of the input 2D image become more complex, and make it difficult for CNN to extract the features with the given structure, which reduces the prediction accuracy. 

\begin{figure}[!h]
    \centering
    \includegraphics[width=3.5 in]{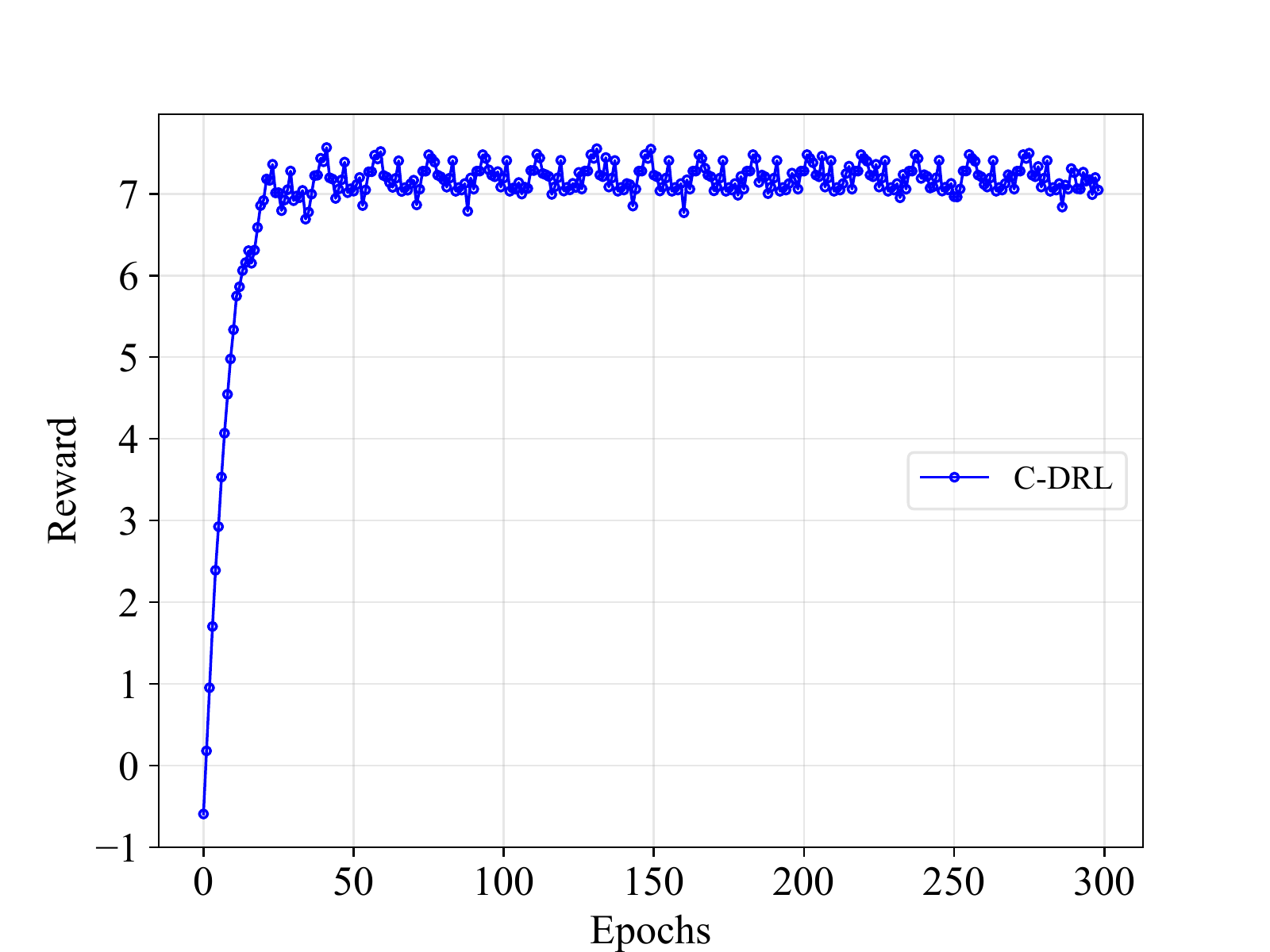}
    \caption{Reward of the MEC-enabled and RIS-assisted THz VR network of each time slot via C-DRL.}
    \label{basic_modules}
\end{figure}

\subsection{RIS Configuration of THz Transmission}
For the downlink THz transmission, we deploy C-DRL to select the proper phase shift of the RIS to reflect the THz signals for the VR users in the NLoS area. For simplicity, we use ``w/ Pred'' to present ``with prediction". In the Genie-aided scheme, the online learning algorithms are trained with known correct actual viewpoint and position of each VR user at each time slot, which is the upper bound of the online learning algorithm and can hardly be achieved in the practical wireless VR systems. To compare with the proposed learning architecture, an exhaustive algorithm is deployed to select the optimal phase shift of the RIS in downlink transmission at each time slot.

\begin{figure*}[ht]
	\centering
	\subfloat[]{\includegraphics[width=3.5in]{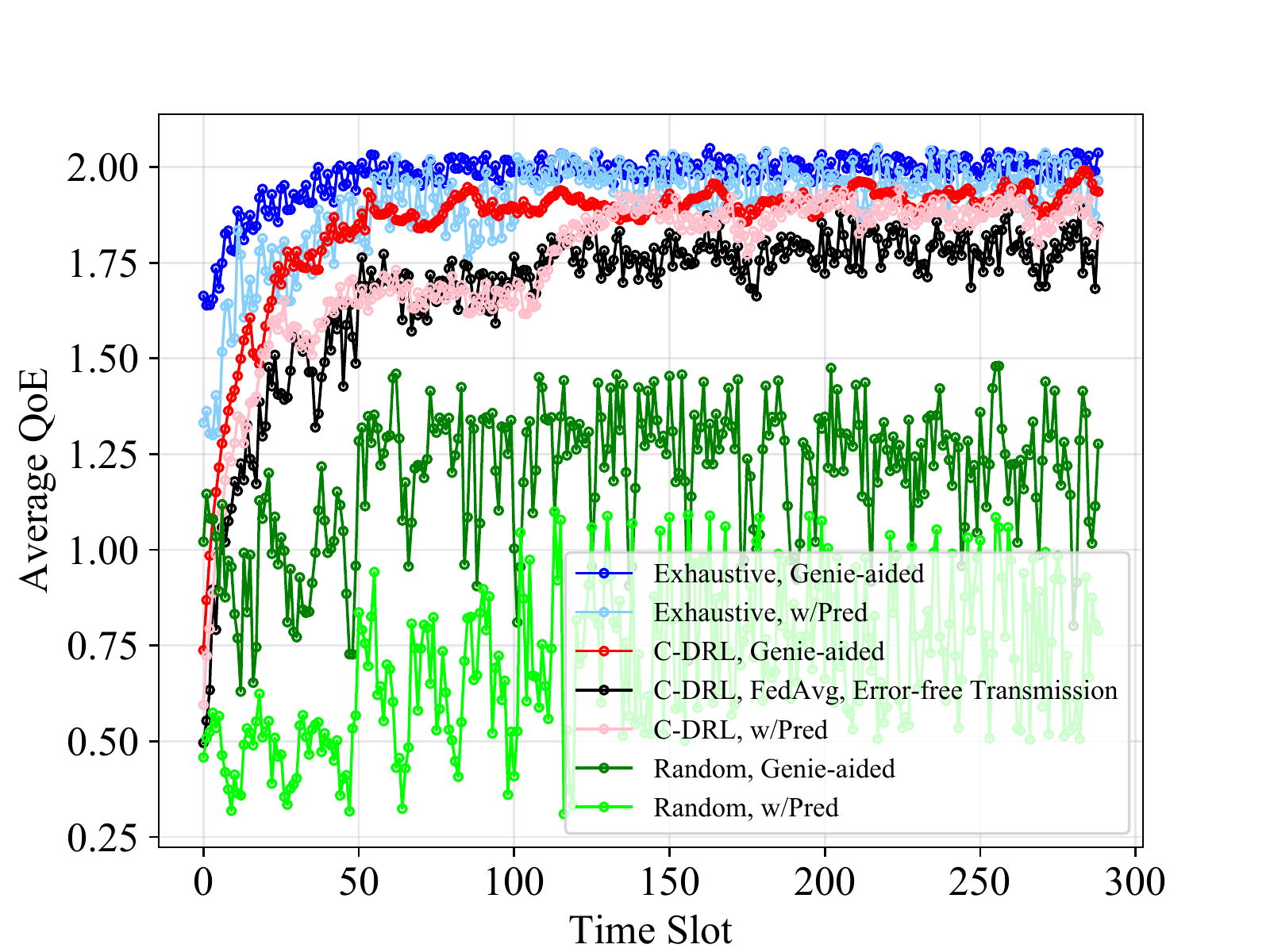}}\label{fig_first_case}
	\subfloat[]{\includegraphics[width=3.5in]{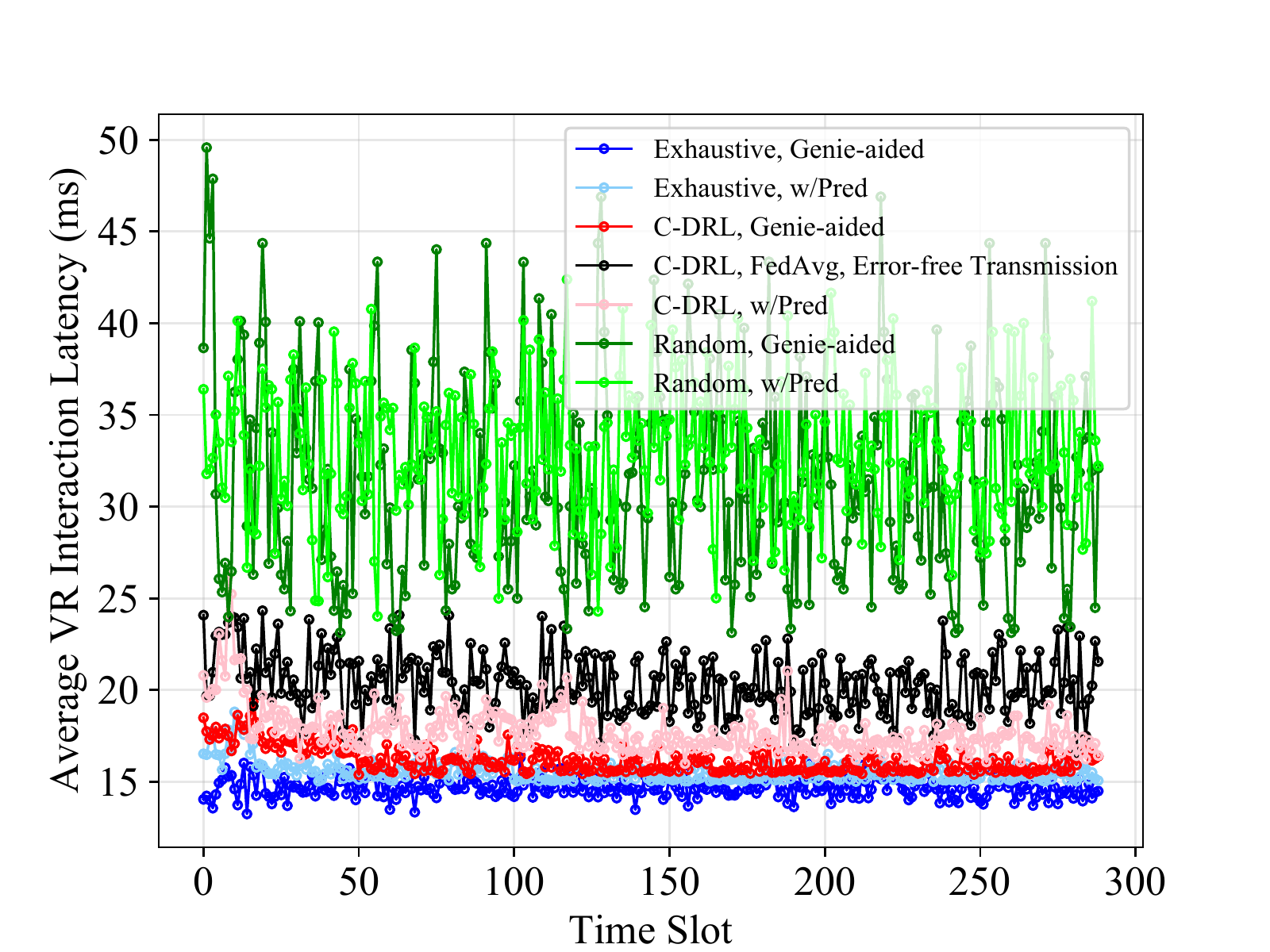}}\label{fig_second_case}
	\caption{(a) Average QoE of the MEC-enabled and RIS-assisted THz VR network of each time slot via C-DRL with viewpoint and LoS/NLoS prediction. (b) Average VR interaction latency of the MEC-enabled and RIS-assisted THz VR network of each time slot via C-DRL with viewpoint and LoS/NLoS prediction, where the VR interaction latency constraint is 20 ms.}
	\label{basic_modules}
\end{figure*}

\begin{figure*}[ht]
	\centering
	\subfloat[]{\includegraphics[width=3.5in]{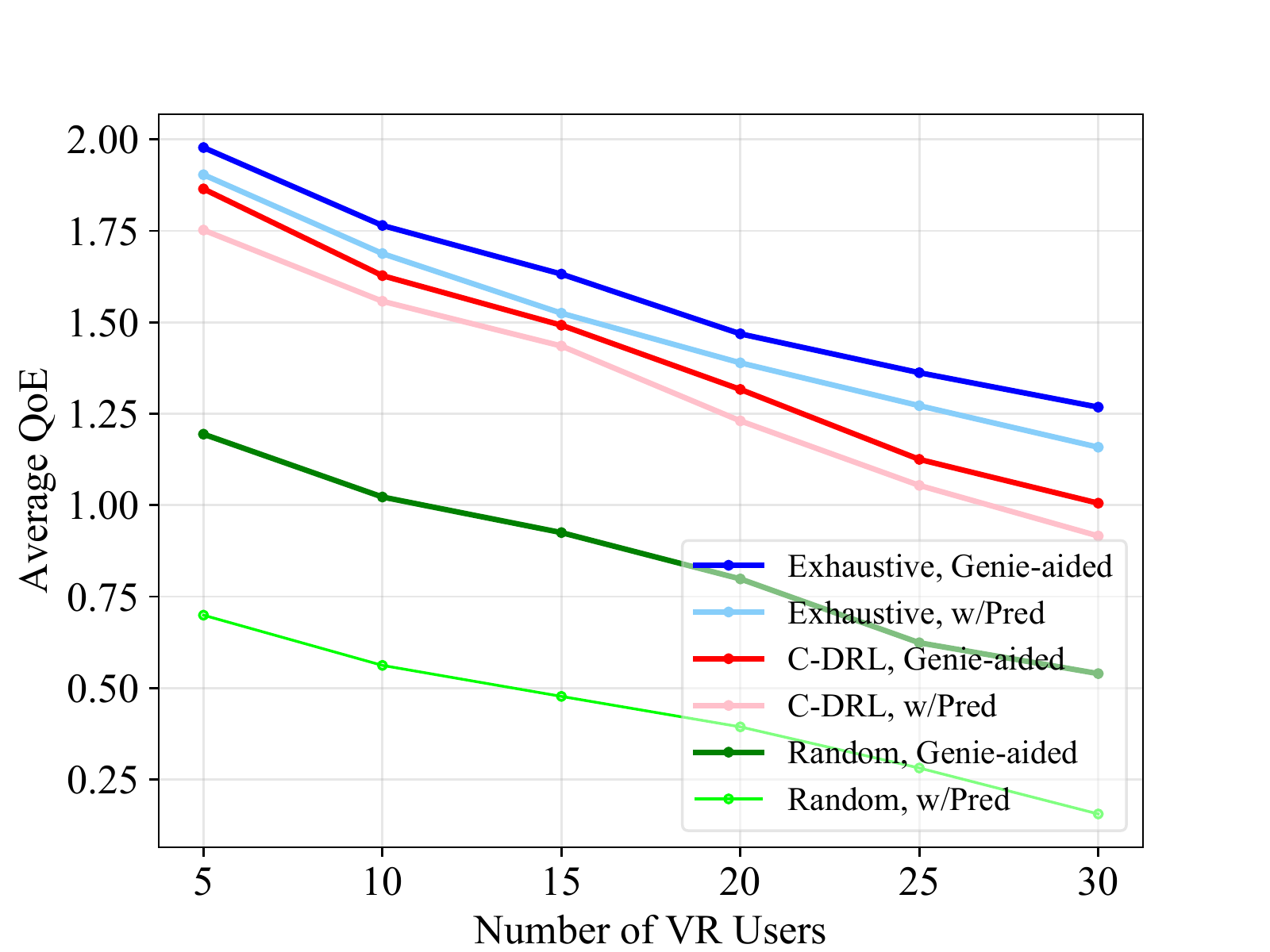}}\label{fig_first_case}
	\subfloat[]{\includegraphics[width=3.5in]{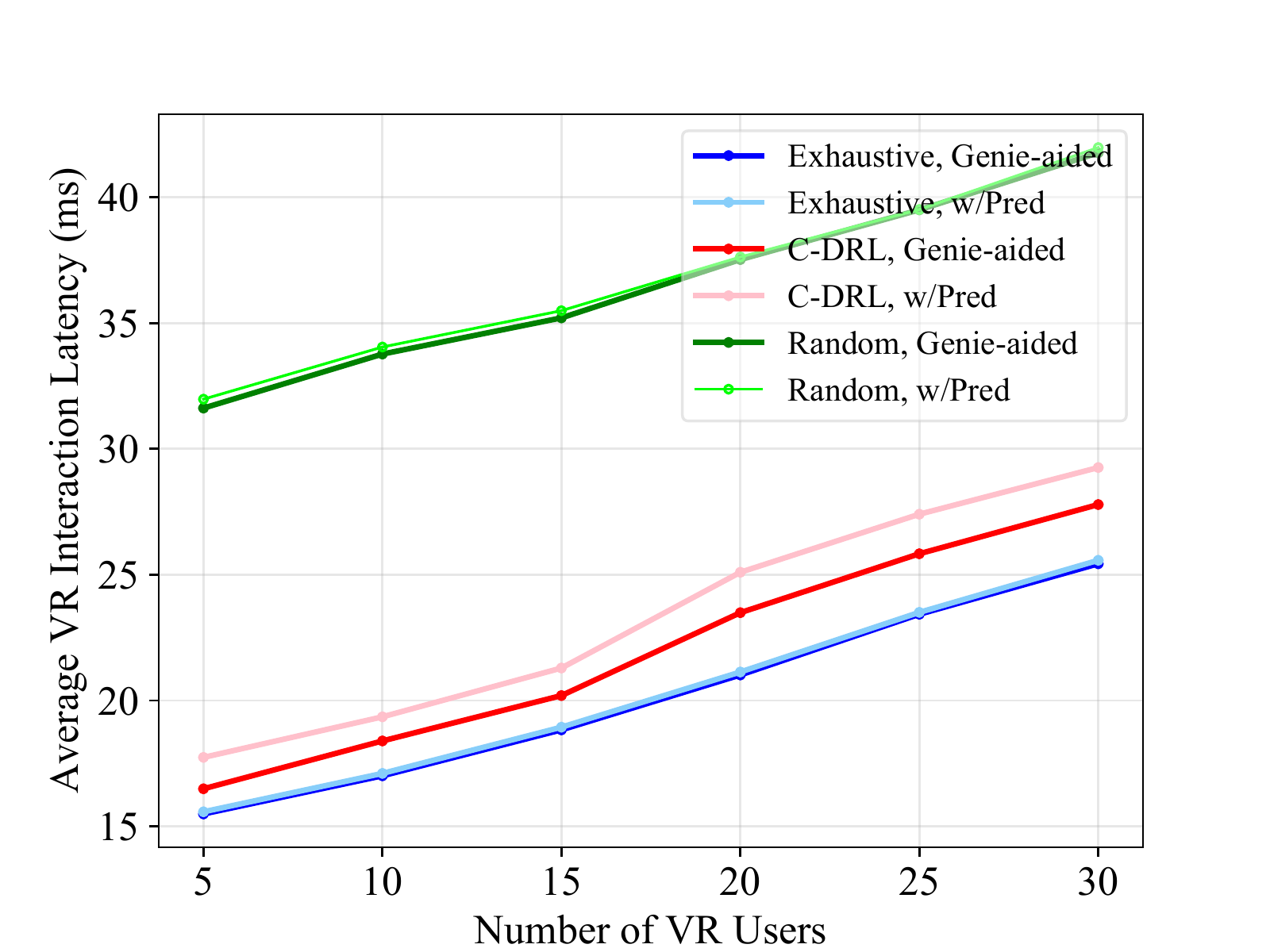}}\label{fig_second_case}
	\caption{(a) Average QoE of the MEC-enabled and RIS-assisted THz VR network via C-DRL with viewpoint and LoS/NLoS prediction with increasing number of VR users. (b) Average VR interaction latency of the MEC-enabled and RIS-assisted THz VR network via C-DRL with viewpoint and LoS/NLoS prediction with increasing number of VR users, where the VR interaction latency constraint is 20 ms.}
	\label{basic_modules}
\end{figure*}

Fig. 11 plots the reward of the MEC-enabled and RIS-assisted THz VR network of each time slot via C-DRL. It can be seen that the C-DRL converges after 50 epochs. Fig. 12 plots the average QoE and the average VR interaction latency of the MEC-enabled and RIS-assisted THz VR network of each time slot via C-DRL with the uplink viewpoint and LoS/NLoS prediction via GRU compared to that via the exhaustive algorithm, respectively. It is observed that at the beginning 150 time slots, the average QoE of C-DRL with prediction scheme is worse than that of the C-DRL with Genie-aided scheme, and both schemes do not violate the VR interaction latency after convergence. This is because in the Genie-aided scheme, the online learning algorithms are directly trained with known correct actual viewpoint and position of each VR user, so that they are capable of better capturing historical trends of viewpoint preference and mobility of the VR user, which can further improve the prediction accuracy.

Interestingly, we notice that after 150 time slots, the gap between the C-DRL with prediction scheme and the exhaustive with prediction scheme is small. This is due to the experience replay mechanism and randomly sampling in C-DRL, which uses the training samples efficiently and smooth the training distribution over the previous behaviours. Importantly, the performance of all the learning-based and exhaustive schemes substantially outperforms the conventional non-learning based scheme, where the reflection coefficients matrix of the RIS is randomly selected. From Fig. 12, we also notice that the performance of the centralized C-DRL with the Genie-aided scheme is better than that of the scheme with FedAvg. This is because in the FedAvg, the learning model needs to be uploaded via uplink transmission for model aggregation, and then the updated global model needs to be transmitted to all VR users through downlink transmission for viewpoint prediction, which leads to extra transmission latency.

\begin{figure*}[ht]
	\centering
	\subfloat[]{\includegraphics[width=3.5in]{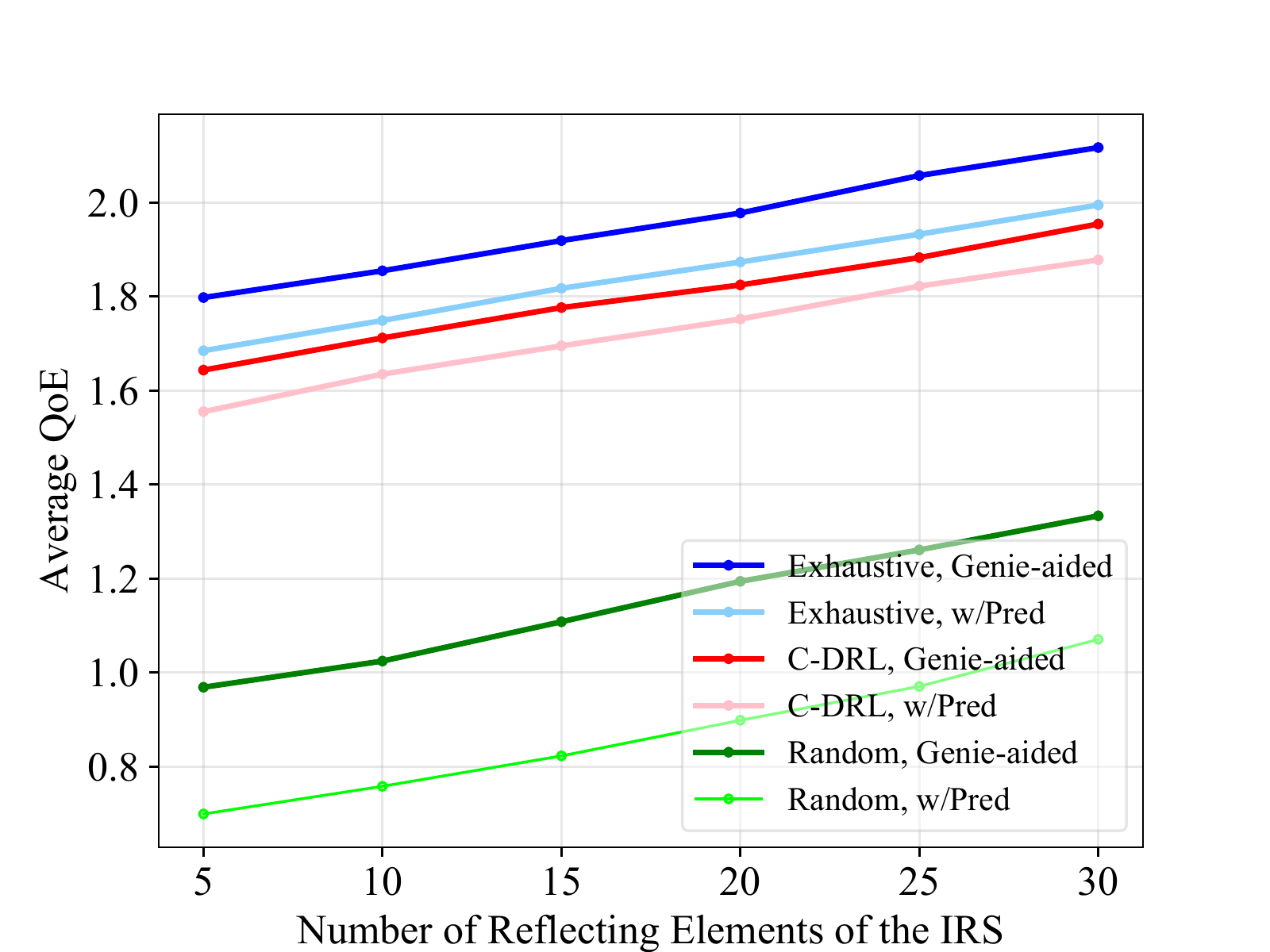}}\label{fig_first_case}
	\subfloat[]{\includegraphics[width=3.5in]{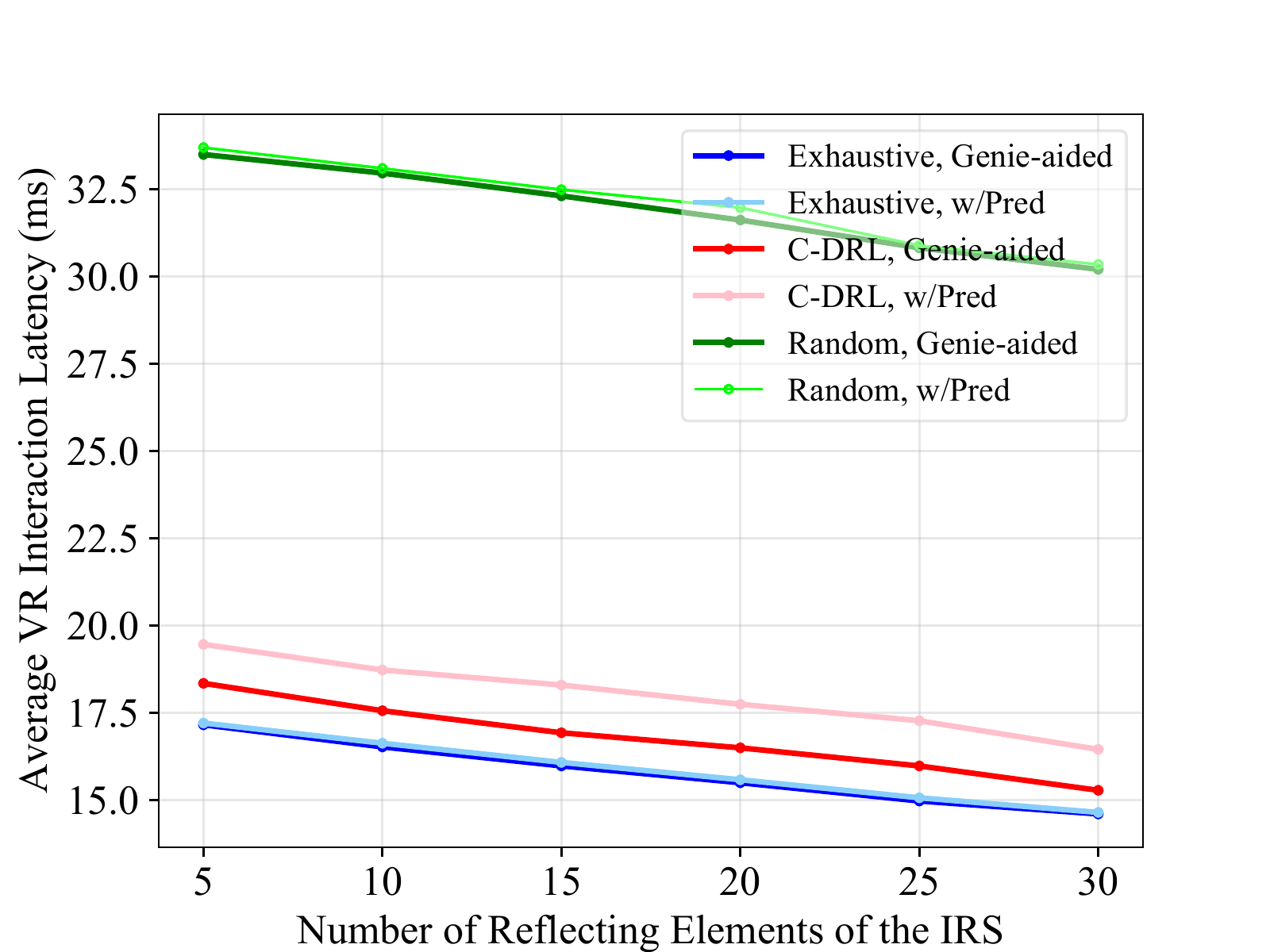}}\label{fig_second_case}
	\caption{(a) Average QoE of the MEC-enabled and RIS-assisted THz VR network via C-DRL with viewpoint and LoS/NLoS prediction with increasing number of reflecting elements of the RIS. (b) Average VR interaction latency of the MEC-enabled and RIS-assisted THz VR network via C-DRL with viewpoint and LoS/NLoS prediction with increasing number of reflecting elements of the RIS, where the VR interaction latency constraint is 20 ms.}
	\label{basic_modules}
\end{figure*}

Fig. 13 plots the average QoE and VR interaction latency of the MEC-enabled and RIS-assisted THz VR network via C-DRL with viewpoint and LoS/NLoS prediction versus the number of VR users compared to that via the exhaustive algorithm, respectively. With the increasing number of VR users, the average QoE of VR users decreases as shown in Fig. 13 (a), whereas the average VR interaction latency increases as shown in Fig. 13 (b). This is due to the fact that with increasing number of VR users, the interference among the THz transmission increases. When the number of the VR user is larger than 15, the gap between the C-DRL and the exhaustive algorithm becomes larger, and the VR interaction latency constraints are violated with increasing number of VR users. This is because the LoS/NLoS prediction accuracy via CNN decreases, which further affects the action selected by the C-DRL.

Fig. 14 plots the average QoE and the average VR interaction latency of the MEC-enabled and RIS-assisted THz VR network via C-DRL with viewpoint and LoS/NLoS prediction versus the number of reflecting elements of the RIS compared to that via the exhaustive algorithm, respectively. With increasing number of reflecting elements of the RIS, the average QoE of VR users increases as shown in Fig. 14 (a), whereas the average VR interaction latency decreases as shown in Fig. 14 (b). This is because as the number of reflecting elements increases, the THz channel gain reflected by the RIS increases \cite{IRS2}, which further increases the THz transmission rate for the VR users in the NLoS area. In addition, the VR interaction latency of the non-learning schemes is not influenced by the predicted LoS/NLoS status via CNN.

\section{Conclusions}
In this paper, a MEC-enabled and RIS-assisted THz VR network was developed to maximize the long-term QoE of real-time interactive VR video streaming in an indoor scenario under VR interaction latency constraints. Specifically, in the uplink, a centralized online GRU algorithm and distributed FedAvg were used to predict the viewpoints of the VR users over time, to determine the corresponding FoV to be rendered at the MEC. An algorithm that integrates online LSTM and CNN was also designed to predict the locations of the VR users and determine the LoS or NLoS statuses in advance. Then, a C-DRL algorithm was developed to select the optimal phase shifts of the reflecting elements of the RIS to compensate for the NLoS loss in THz transmission. Simulation results have shown that our proposed ensemble learning architecture with online GRU, online LSTM, CNN, and C-DRL algorithms substantially improved the long-term QoE, while satisfying the VR interaction latency constraint, and the QoE performance of our proposed learning architecture was near-optimal compared to the exhaustive algorithm.

%





\ifCLASSOPTIONcaptionsoff
  \newpage
\fi





\bibliographystyle{IEEEtran}
\bibliography{IEEEabrv,Ref,ReferencesMP}
%
%

\end{document}